\documentclass[aps,pra,notitlepage,reprint,superscriptaddress,nofootinbib]{revtex4-1}

\usepackage{graphicx}
\usepackage{dcolumn}
\usepackage{bm}
\usepackage{amsmath}
\usepackage{upgreek}
\usepackage[colorlinks,linkcolor=red,citecolor=blue,urlcolor=red]{hyperref}
\usepackage[utf8]{inputenc}
\usepackage[T1]{fontenc}
\usepackage{mathptmx}
\usepackage{ulem}
\usepackage[colorinlistoftodos]{todonotes}
\renewcommand{\t}[1]{\mathrm{#1}}
\begin{document}

\title{Nanoscale torsional dissipation dilution for quantum experiments and precision measurement}

\author{J. R. Pratt}
\affiliation{National Institute of Standards and Technology, 100 Bureau Drive, Gaithersburg, MD 20899}

\author{A. R. Agrawal}
\author{C. A. Condos}
\author{C. M. Pluchar}
\affiliation{Wyant College of Optical Sciences, University of Arizona, Tucson, AZ 85721, USA}

\author{S. Schlamminger}%
\affiliation{National Institute of Standards and Technology, 100 Bureau Drive, Gaithersburg, MD 20899}

\author{D. J. Wilson}
\affiliation{Wyant College of Optical Sciences, University of Arizona, Tucson, AZ 85721, USA}

\date{\today}

\begin{abstract}
	
We show that torsion resonators can experience massive dissipation dilution due to nanoscale strain, and draw a connection to a century-old theory from the torsion balance community which suggests that a simple torsion ribbon is naturally soft-clamped. By disrupting a commonly held belief in the nanomechanics community, our findings invite a rethinking of strategies towards quantum experiments and precision measurement with nanomechanical resonators.  For example, we revisit the optical lever technique for monitoring displacement, and find that the rotation of a strained nanobeam can be resolved with an imprecision smaller than the zero-point motion of its fundamental torsional mode, without the use of a cavity or interferometric stability.  We also find that a strained torsion ribbon can be mass-loaded without changing its $Q$ factor.  We use this strategy to engineer a chip-scale torsion balance whose resonance frequency is sensitive to micro-$g$ fluctuations of the local gravitational field.  Enabling both these advances is the fabrication of high-stress Si$_3$N$_4$ nanobeams with width-to-thickness ratios of $10^4$ and the recognition that their torsional modes have $Q$ factors scaling as their width-to-thickness ratio squared, yielding $Q$ factors as high as \mbox{$10^8$ and $Q$-frequency products as high as $10^{13}$ Hz.}

\end{abstract}

\maketitle

Recent years have seen the emergence of a new class of ultra-high-$Q$ nanomechanical resonators fashioned from strained thin films \cite{sementilli2021nanomechanical}. The mechanism behind their performance is dissipation dilution, 
an effect whereby elastic body is subjected to a conservative stress field, increasing its stiffness without adding loss \cite{Gonzales1994, Federov2019_Generalized}. Access to extreme dimensions and stresses at the nanoscale has enabled dilution factors (the ratio of final to initial $Q$) in excess of $10^5$, yielding $Q$ factors in excess of $10^9$ for devices made of amorphous glass and $Q$-frequency products exceeding $10^{15}$ Hz using "soft-clamping" \cite{tsaturyan2017ultracoherent,ghadimi2018elastic}. Attractive features of these devices include attonewton force sensitivies \cite{reinhardt2016ultralow}, thermal coherence times of milliseconds \cite{tsaturyan2017ultracoherent}, and zero-point displacement amplitudes in excess of picometers \cite{ghadimi2018elastic}, spurring proposals from room temperature quantum experiments \cite{norte2016mechanical} to ultra-fast force microscopy \cite{halg2021membrane}.

Despite rapid innovation, a key limitation of dissipation dilution is its restriction to transverse flexural modes, a consequence of its reliance on nonlinear stress-strain coupling.  It has been formally shown \cite{Federov2019_Generalized} that breathing modes, such as the longitudinal modes of a cylinder, cannot be diluted by a uniform strain field, ruling out the application of dissipation dilution to a large class of metrologically important mechanical devices, such as the mirrors used in precision optical cavities and gravitational wave interferometers. It is also commonly held that torsional modes of nanostructures are not diluted by strain, despite the prevalence of tensioned microsuspensions in macroscopic torsion pendula \cite{Quinn2014} and the historical use of micro-torsion resonators to study mechanical dissipation \cite{Bishop1985}.

Here we show that torsional resonators can experience massive dissipation dilution due to nanoscale strain, and draw a connection to a century-old theory from the torsion balance community which suggests that a simple torsion ribbon is naturally soft-clamped. By disrupting a commonly held belief in the nanomechanics community, our findings invite a rethinking of strategies towards quantum experiments and precision measurement with nanomechanical resonators.  For example, we revisit the optical lever technique for monitoring displacement \cite{putman1992}, and find that the rotation of a strained nanobeam can be resolved with an imprecision smaller than the zero-point motion of its fundamental torsional mode, without the use of a cavity or interferometric stabilty.  We also find that a strained torsion ribbon can be mass-loaded without changing its $Q$ factor.  We use this strategy to engineer a chip-scale torsion balance whose resonance frequency is sensitive to micro-$g$ fluctuations of the local gravitational field.  Enabling both these advances is the fabrication of high-stress Si$_3$N$_4$ nanobeams with width-to-thickness ratios of $10^4$ and the recognition that their torsional modes have $Q$ factors scaling as the width-to-thickness ratio squared, yielding $Q$ factors as high as $10^8$ and $Q$-frequency products as high as $10^{13}$ Hz.

\begin{figure}[b!]
\vspace{-5mm}
	\includegraphics[width=0.68\columnwidth]{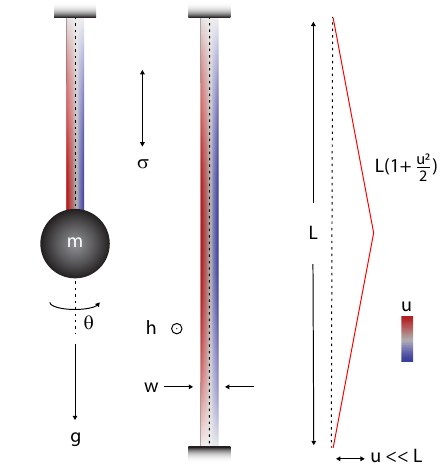}
	\caption{\textbf{Dissipation dilution of a torsion strip.} (a) A torsion pendulum formed by suspending a rigid mass $m$ from beam-like torsion fiber. Gravity loads the fiber into tension $T = mg$. (b) A beam of width $w$ and thickness $h$ loaded under tensile stress $\sigma = T/wh$. \textcolor{black}{(c) Bifilar model of torsion mode \cite{SI}: transverse displacement $u$ preserves the beam length $L$ to first order, leading to dissipation dilution.}}
	\label{fig1}
\vspace{-2mm}
\end{figure}

\begin{figure*}[t!]
\vspace{-3mm}
	\includegraphics[width=2\columnwidth]{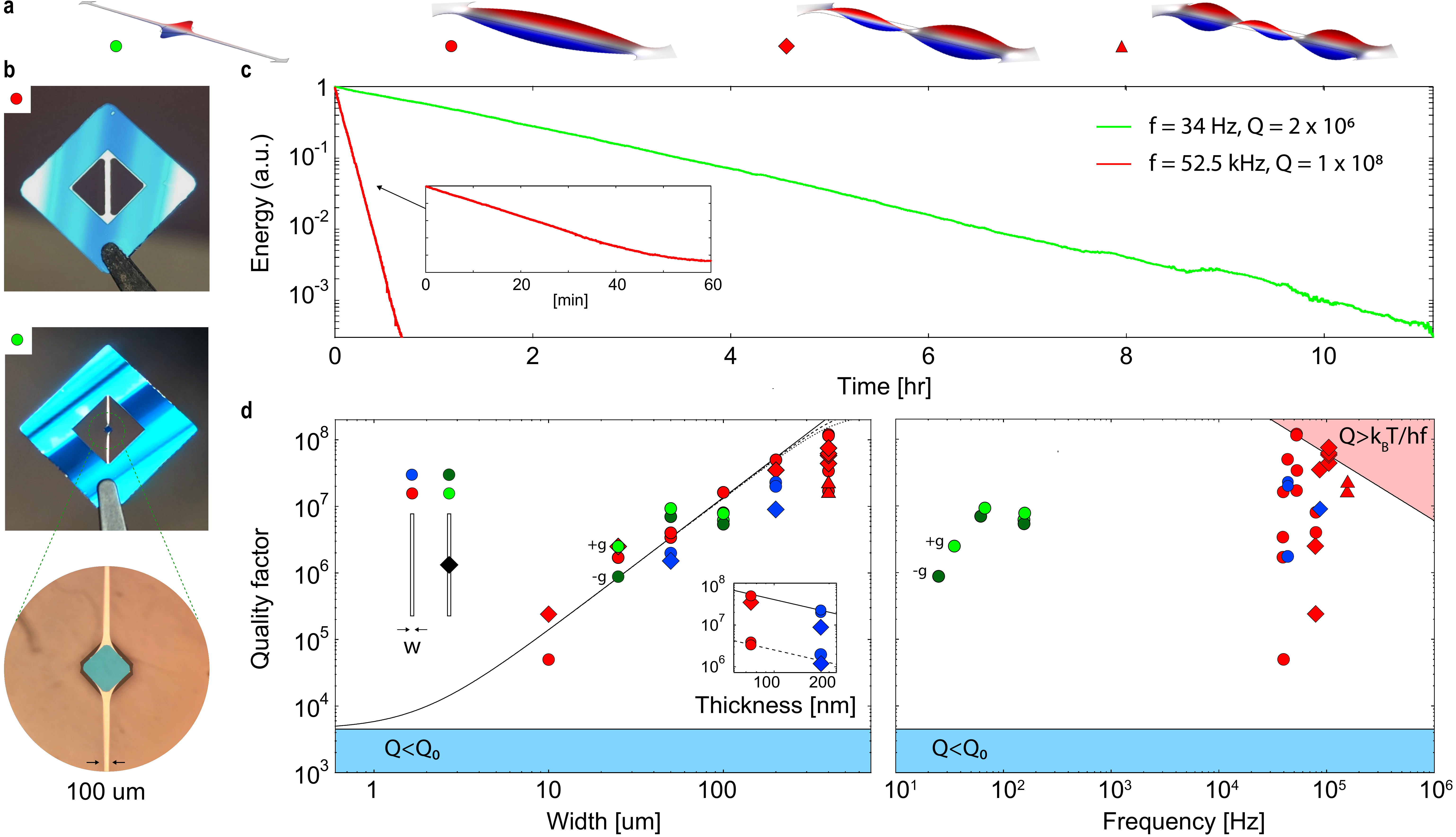}
	\caption{\textbf{Ultra-low-loss nanomechanical torsion resonators.} (a) Finite element models of torsion modes for a nanobeam with (green) and without (red) mass loading.  (b) Photos of representative devices: (top) a 400 um wide, 7 mm long, 75 nm thick beam; (middle) a 25 um wide, 7 mm long, 75 nm thick beam loaded with a 100 um thick Si paddle; and (bottom) a micrograph of the paddle. (c) Ringdowns of the fundamental mode of an unloaded, 400 um wide beam (red) and a loaded, 25 um wide beam (green). (d) Compilation of $Q$ versus beam width (left) and resonance frequency (right).  Red and blue points correspond to thicknesses of 75 and 180 nm, respectively. Dark and light green points correspond to inverted ($-g$) and non-inverted ($+g$) orientation. Solid dashed, and dotted lines correspond to the lumped mass model, finite element simulations, and a gas damping model with $Q_\t{gas}=1.1\times10^9$ \cite{SI}. \mbox{Inset: $Q$ versus thickness for 200 $\upmu$m (solid) and 50 $\upmu$m (dashed) wide beams.}}
	\label{fig2}
\vspace{-3mm}
\end{figure*}

\textit{Bifilar suspensions and torsional dissipation dilution-} The history of torsional dissipation dilution can be traced back to the theory of bifilar suspensions by Buckley in 1914 \cite{Buckley1914}.  In brief, loading a torsion strip with a massive plumb bob increases its stiffness without adding loss \cite{Quinn2014}.  This is because as the strip twists, it lifts the bob and does work against a conservative gravitational field.  Vibrating in its fundamental mode, the $Q$ factor of the loaded torsion strip scales as the ratio of the intrinsic (elastic) stiffness to the loaded stiffness
\begin{equation}\label{eq1}
\frac{Q}{Q_0}=1+\frac{k_\sigma}{k_E}\approx \frac{\sigma}{2E}\left(\frac{w}{h}\right)^2
\end{equation}
where $\sigma = T/wh$ is the tensile stress produced by mass loading ($T=mg$) a ribbon of width $w$ and thickness $h$ \cite{SI}.

Nanoscale dissipation dilution theory \cite{Federov2019_Generalized} holds that any form of static tensile stress, such as residual stress in thin films, can give rise to dissipation dilution. Buckley's theory can therefore be generalized to the idea that \textit{placing a ribbon under tensile stress increases its stiffness without adding loss}.  Eq. \ref{eq1} should thus hold for a nanobeam released from a thin film with biaxial stress $\sigma$.  Remarkably, Eq. \ref{eq1} is also known to be the ideal ``soft-clamped" dissipation dilution factor for a thin beam vibrating in its fundamental flexural mode (with width replaced by half length) \cite{ghadimi2018elastic}.  In the appendix, we provide a continuum mechanics model that supports this claim, \mbox{which we shall now proceed to investigate experimentally.}

\textit{Torsional dissipation dilution of Si$_3$N$_4$ nanobeams-} To investigate Eq. \ref{eq1}, we fabricated a set of high-stress Si$_3$N$_4$ nanobeams with aspect ratios varying from $w/h\sim 10^2-10^4$ \cite{SI} (Fig. \ref{fig2}a).  Devices were housed in a high vacuum chamber to minimize gas damping and ringdowns were performed using optical lever measurements in conjunction with radiation pressure driving \cite{SI}.  Ringdown-inferred $Q$ factors were then compiled for flexural and torsional modes up to third order, as shown in Fig. \ref{fig2}d and in the appendix.

Considering first the hypothesis that $Q\sim (w/h)^2$, in Fig. 2d we compare $Q$ factors of beams with widths from $10\,\upmu\t{m}$ to $400\,\upmu\t{m}$ to Eq. \ref{eq1} and a finite element simulation accounting for filleting \cite{SI}. For both models we assume $\sigma = 0.85\,\t{GPa}$, $E = 250\,\t{GPa}$, $h=75$ nm, and $Q_0 = 6000\times h/(100\,\t{nm})$ (an established surface loss model for Si$_3$N$_4$ thin film resonators \cite{villanueva2014evidence}).  We observe quantitative agreement with both models up to a width of $100\,\upmu\t{m}$, beyond which $Q$ begins to drop, consistent with simulated buckling instabilities \cite{kudrolli2018tension,SI} and residual gas damping at the level of $Q_\t{gas}\approx 10^9$ (dotted line) \cite{SI}.  At widths smaller than 100 $\upmu$m, we also observe a slightly higher $Q$ than predicted.  We conjecture that this may be due to an underestimate of $Q_0$, since the surface loss model in \cite{villanueva2014evidence} is inferred from a study of flexural rather than torsional modes.

Further support for Eq. \ref{eq1} was obtained by inspecting higher order modes and by varying the thickness of several resonators. The continuum model derived in the appendix predicts that $Q_n/Q_0$ is independent of mode order $n$ for acoustic wavelengths $L/n\gg w$, in consistency with our observations that $\{Q_n\}$ is bound by Eq. \ref{eq1}.  We also observed that the fundamental torsional mode had consistently high $Q$.  This is contrary to the typical behavior of flexural modes of Si$_3$N$_4$ beams and membranes \cite{villanueva2014evidence}, and suggests that torsional modes may be more resistant to acoustic radiation (``mounting'') loss \cite{cole2011phonon}. 

Finally, we fabricated several beams with thickness 180 nm, and observed that the $Q$ of torsional modes scaled inversely with thickness (Fig. \ref{fig2}d inset). By contrast, as shown in the appendix \cite{SI}, we observed that the $Q$ of flexural modes were roughly independent of width and thickness.  These different scalings are tell-tale signatures of "soft-clamping" ($Q/Q_0\propto w^2/h^2$) and "hard-clamping" ($Q/Q_0\propto w/h$) \cite{schmid2011damping}, respectively, in the presence of surface loss ($Q_0\propto h$).

\textit{Quantum-limited deflectometry of a nanoribbon-} Nanomechanical resonators have been probed at the quantum limit using cavity-enhanced interferometry \cite{purdy2013observation, mason2019continuous}.  \textcolor{black}{In principle, however, neither a cavity nor interferometry is necessary}, provided that \textcolor{black}{the measurement is optimally efficient} \cite{clerk2010introduction}.  In our study of Si$_3$N$_4$ nanobeams, we have found that the imprecision  of optical lever measurements $S_\theta^\t{imp}$ (here expressed as a power spectral density) can be reduced to below the zero-point angular displacement of the beam's fundamental torsion mode $S_\theta^\t{ZP}$, satisfying a basic requirement for displacement measurement at the Standard Quantum Limit ($S_\theta^\t{imp}+S_\theta^\t{ba} = S_\theta^\t{ZP}$, where $S_\theta^\t{ba}$ is the displacement produced by back-action torque $S_{\tau}^\t{BA}\ge\hbar/S_\theta^\t{ZP}$) \cite{teufel2009nanomechanical,enomoto_standard_2016}.  To our knowledge, this represents the first non-interferometric displacement measurement with an imprecision below that at the Standard Quantum Limit. Combined with robustness to \textcolor{black}{misalignment and technical noise} \cite{fukuma_development_2005}, access to quantum-limited deflection measurements signals the potential for a new generation of quantum optomechanics experiments employing cryogenic or even room temperature nanotorsion resonators coupled to optical levers.

\begin{figure}[t!]
	\includegraphics[width=0.9\columnwidth]{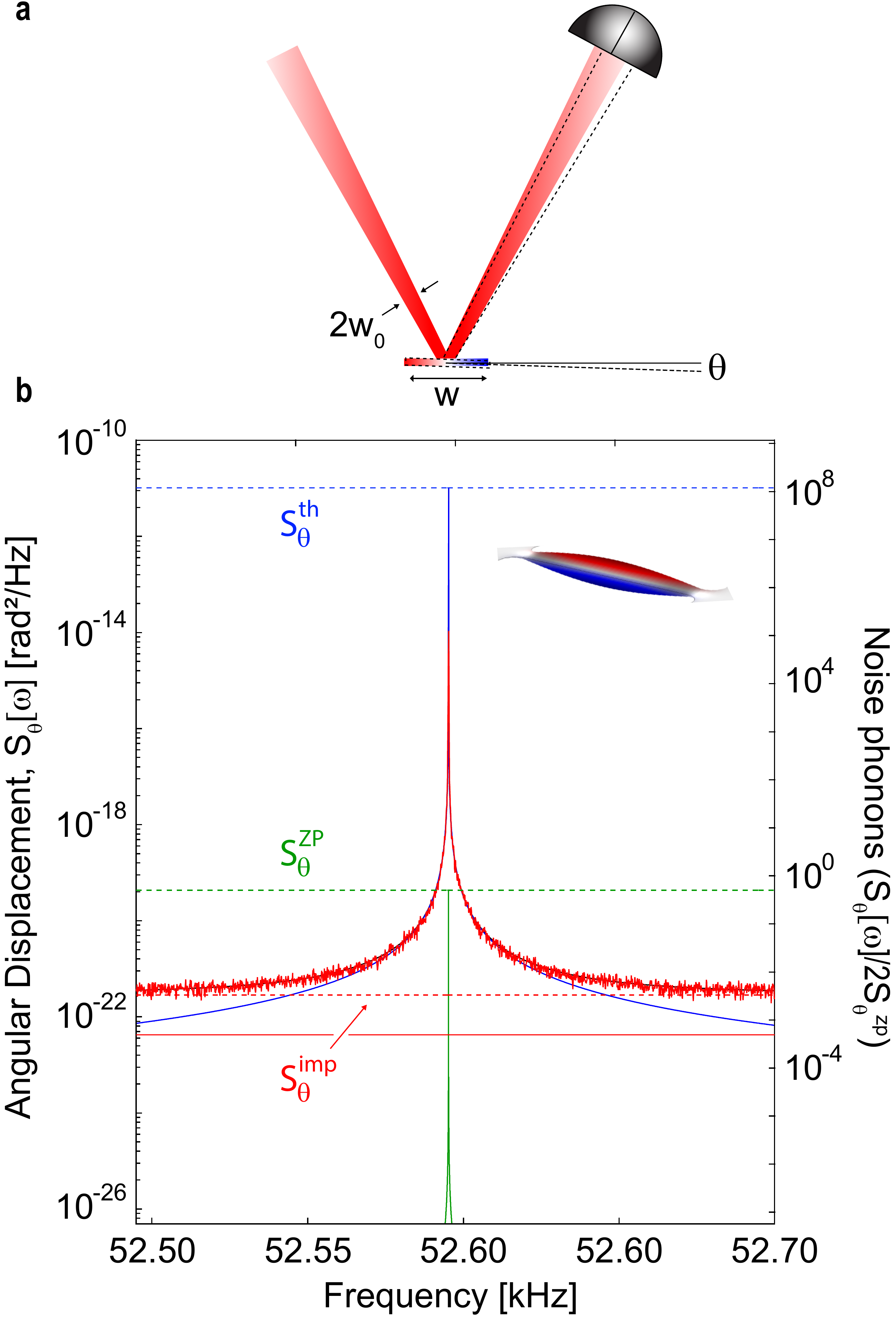}
	\caption{\textbf{Optical lever measurement with an imprecision below that at the Standard Quantum Limit.} (a) Schematic of the optical lever technique: A Gaussian beam with a waist of $w_0$ is reflected off of torsion beam of width $w$.  Beam deflection $\theta$ is monitored using a split photodiode. (b) Displacement of a $w = 400\;\upmu$m beam measured using a $w_0 \approx 200\,\upmu$m, $P=4$ mW optical lever. Dotted lines are inferred from fitting and solid red line is estimated using Eq. \ref{eq2}.}
	\label{fig3}
	\vspace{-5mm}
\end{figure}

To explore the potential for torsional quantum optomechanics, we have revisited the optical lever technique with an eye to maximizing the ratio $S_\theta^\t{ZP}/S_\theta^\t{imp}$ for a torsion beam. As shown in Fig. \ref{fig3}a, the ``lever'' is formed by reflecting a laser field off the beam and monitoring its deflection on a split photodiode. In the far-field, angular displacement of the beam $\theta$ can be resolved with a shot-noise-limited resolution of
\begin{equation}
\label{eq2}
S_\theta^\t{imp}\gtrsim\frac{1}{w_0^2}\frac{\hbar c \lambda}{8P}
\end{equation}
where $P$ is the reflected power and $w_0$ is the spot size of the laser field.  Comparing to the zero-point displacement spectral density of the beam's
fundamental torsion mode
\begin{equation}
\label{eq3}
S_\theta^\t{ZP}=\frac{1}{w^2}\frac{8\hbar Q_1}{m_1\omega_\t{1}^2},
\end{equation}
(effective mass $m_1 = \rho h w L/6$), we find that maximum ``leverage'' is achieved by matching the width of the optical and mechanical beams ($w_0\approx w/2$), giving access to a favorable scaling $S_\theta^\t{ZP}/S_\theta^\t{imp}\propto Q_0w/h^3$ due to dissipation dilution \cite{SI}.

An optical lever measurement with an imprecision below that at the Standard Quantum Limit ($S_\theta^\t{imp}\le S_\theta^\t{ZP}/2$) is shown in Fig. \ref{fig3}. The measurement was made by reflecting a $w_0\approx 200\;\upmu$m wide optical field from a $400$ $\,\upmu\t{m}$ wide, $75\,\t{nm}$ thick nanobeam and detecting the 4 mW reflected field at a distance of $0.4$ meters. Near the beam's fundamental torsion resonance $\omega_1\approx 52.5\,\t{kHz}$, 
the photocurrent spectrum is dominated by thermal noise with a peak magnitude of $S_\theta^\t{th}=2n_\t{th}S_\theta^\t{ZP}$, 
well in excess of zero point motion due to the large thermal mode occupation, $n_\t{th} = k_\t{B} T/\hbar\omega_1 = 1.2\times 10^8$. Fitting the noise peak to a Lorentzian $S_\theta^\t{imp}+S_\theta^\t{th}/(1+4Q_1^2(\omega-\omega_1)^2/\omega_1^2)$, where $Q_1 = 8.5\times10^7$ is determined separately by ringdown, we infer that the measurement imprecision $S_\theta^\t{imp}$ is factor of \textcolor{black}{140 below the zero-point spectral density $S_\theta^\t{ZP}$ and a factor of 14 above the ideal value implied by Eq. \ref{eq2}.} While this inference is independent of the modal mass ($m\approx 50\,\t{ng}$), we speculate that the heirarchy $S_\theta^\t{ZP}\gg S_\theta^\t{imp}$ is aided by the large magnitude of the zero-point motion $S_\theta^\t{ZP}\approx (0.5\,\t{nrad}/\sqrt{\t{Hz}})^2$ and the immunity of deflectometry to various forms technical noise \cite{SI}.

%

\begin{figure}[ht!]
	\vspace{-2mm}
	\includegraphics[width=0.87\columnwidth]{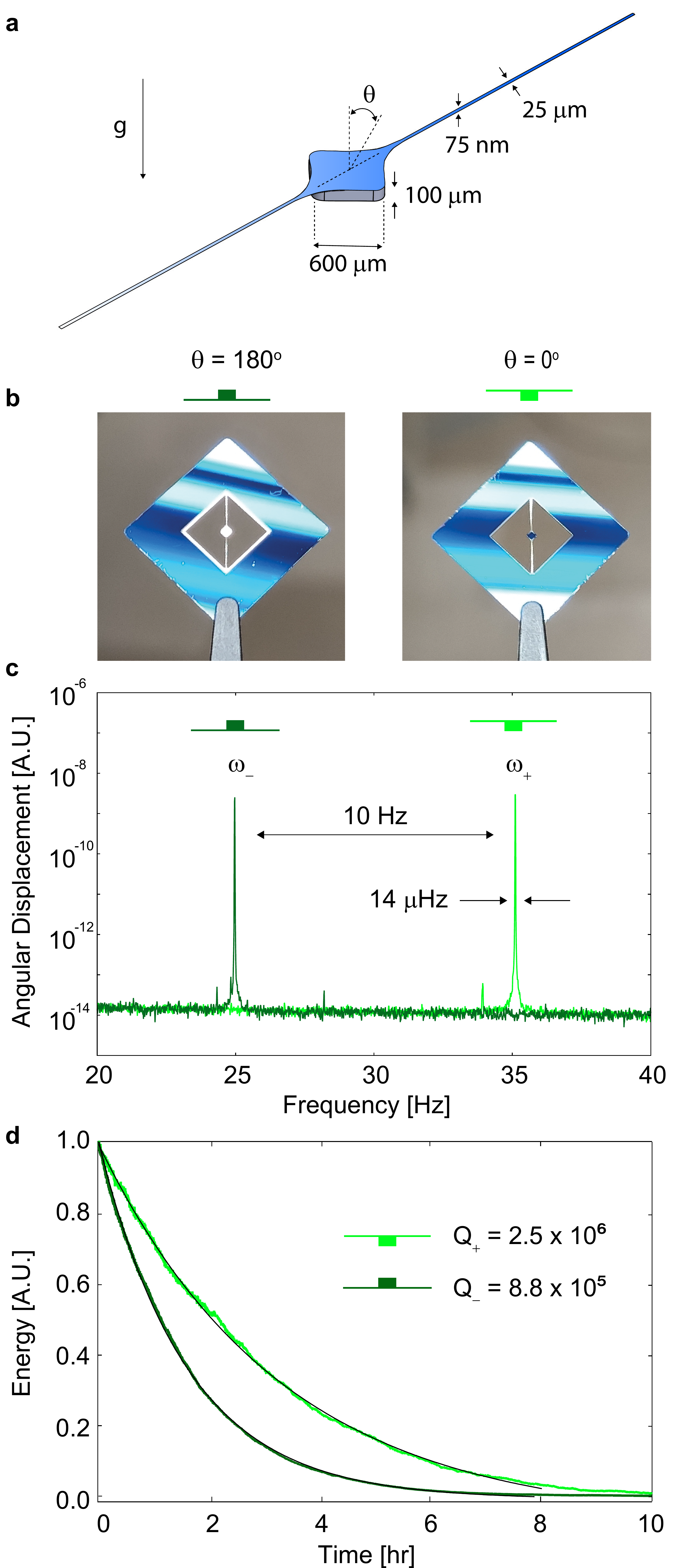}
	\caption{\textbf{A microtorsion pendulum for weak gravity measurement}. (a) Geometry of the torsion pendulum, produced by mass-loading a Si$_3$N$_4$ nanoribbon (blue) with a rigid Si pad (gray).  (b) Photographs of the device in non-inverted and inverted configuration. (c) Free-running deflection spectrum of the pendulum in non-inverted and inverted configuration.  (d) Ringdowns of the fundamental torsional mode in non-inverted and inverted configuration.  }
	\label{fig4}
	\vspace{-5mm}
\end{figure}

\textit{A nanoribbon-based torsion gravimeter-} A powerful feature of nanobeams is their ability to be mass-loaded without reducing their torsional $Q$ factor. The concomitant reduction in resonance frequency and damping rate is useful for a variety of inertial sensing applications.  For instance, by asymmetrically loading a beam (Fig. \ref{fig4}a), 
a torsion pendulum is created whose resonance frequency depends on the local acceleration of gravity. The gravity-equivalent damping rate of the pendulum gives a measure of its sensitivity as a gravimeter \cite{SI}:
\begin{equation}\label{eq4}
\Delta g_\t{min} = \frac{4g_0}{Q_+}\frac{\omega_+^2}{\omega_+^2-\omega_-^2}= \frac{2g_0}{Q_0}\frac{k_E}{k_g}%
\end{equation}
where $g_0 = 9.8\,\t{m}/\t{s}^2$ is the standard gravity of earth and
\begin{subequations}\label{eq5}\begin{align}
\omega_\pm & =\sqrt{(k_E+k_\sigma\pm k_g)/I}\\
Q_\pm&=Q_0\left(1+\frac{k_\sigma\pm k_g}{k_E}\right)= Q_\mp \left(\frac{\omega_\mp}{\omega_\pm}\right)^2
\end{align}\end{subequations}
are the resonance frequency and $Q$ factor of the pendulum in the inverted ($-$) and noninverted ($+$) configuration. 

We have realized a passively stable, chip-scale torsion pendulum sensitive to micro-$g$ gravity fluctuations ($\Delta g_\t{min}\sim g_0/10^6$) by suspending a rigid Si mass from Si$_3$N$_4$ nanoribbon.  To understand this advance, it is important to first note that Eq. \ref{eq4} implies that the sensitivity of a pendulum gravimeter is \textit{independent} of tensile stress and is maximized by overwhelming the elastic stiffness of the torsion fiber $k_E$ with gravitational stiffness due to mass-loading $k_\t{g}$. 
Historically, this insight has driven the pursuit of unstressed torsion fibers \cite{LaCoste}, 
culminating most recently in fibers made of 2D materials \cite{cong2021chip}; however, a  limitation of this approach is that the bias point of the torsion fiber becomes unstable for $k_\t{g}>k_E$, necessitating the use of anti-springs to balance the static load \cite{Middlemiss2016,Tang2019}, adding complexity, and mechanical loss.  Provided it does not add dissipation (Eq. \ref{eq5}b), tensioning the torsion fiber restores bias stability as long as $k_\sigma+k_E\gtrsim k_\t{g}$, giving access to large sensitivity enhancement ($k_g/k_E\gg 1$) for ribbon-like torsion fibers made of a sufficiently strained material that $k_\sigma\gg k_E$.

As shown in Fig. \ref{fig4},  we formed our pendulum by suspending a $100\;\upmu\t{m}$ thick, $600\times 600\; \upmu\t{m}^2$ wide Si pad beneath a 75 nm thick, $25\,\upmu\t{m}$ wide  Si$_3$N$_4$ beam by under-etching a central defect \cite{SI}. In-vacuum measurements revealed a 1000-fold drop in the fundamental torsion resonance frequency of the beam from 40 kHz to 34 Hz, corresponding to a million-fold increase in moment of inertia. Despite this substantial mass-loading, ringdown measurements revealed an \textit{increased} quality factor of $Q_+\approx 2.5\times10^6$ relative to the unloaded beam ($Q_0\approx 1.5\times 10^6$), consistent with gravitational dissipation dilution (Eq. \ref{eq5}b) with a stiffness hierarchy of $k_\sigma\approx 2k_g \approx 200 k_E$.  In principle, the resonance frequency of this pendulum should be sensitive to gravity at the level of $\Delta g_\t{min} = 2\times 10^{-6} g_0$.

To explore it's potential as a gravimeter, we inverted the pendulum and remeasured its resonance frequency and $Q$ factor. As shown in Fig. \ref{fig4}c-d, we observed a 10 Hz drop in resonance frequency ($\omega_-/\omega_+ = 0.71$) and a three-fold drop in $Q$ factor ($Q_-/Q_+ = 0.35$), in qualitative agreement with Eq. \ref{eq5}b.  In practice the performance of a pendulum gravimeter depends on its frequency stability.  Our preliminary attempts to track the resonance frequency of the pendulum in its non-inverted configuration  yielded a minimum Allan deviation of $\sigma_{\delta\omega_+/\omega_+}\approx 2\times10^{-6}$ at 600 seconds (see appendix \cite{SI}), corresponding to a gravity uncertainty of $\sigma_{\delta g/g}\approx 8\times10^{-6}g_0$. 

Finally, we note that similar sensitivity can be obtained with wider, stiffer torsion ribbons possessing higher $Q$ factors. As shown in Fig. \ref{fig2}c, micro-torsion pendula with ribbon widths varied between 25 and 100 um were investigated.  The $Q$ factor and resonance frequency of these devices was found to scale quadratically with width, in agreement with the lumped mass model.  For a pendulum based on a $50\,\upmu$m wide ribbon, we observed a loaded quality factor as high as $Q_+\approx 1.0\times 10^7$ and an inverted frequency ratio of $\omega_-/\omega_+ = 0.92$, corresponding to a gravity-equivalent damping rate of $\Delta g_\t{min} = 3\times 10^{-6} g_0$.

\textit{Summary and outlook- } We have shown that torsion modes of nanostructures can experience massive dissipation dilution due to thin film stress, paving the way for a new class of ultra-high-$Q$ nanomechanical resonators with applications to quantum experiments and precision measurement.  Specifically, we studied high-stress Si$_3$N$_4$ nanobeams with width-to-thickness ratios as large as $10^4$, and found that their torsion modes have $Q$ factors that scale as the width-to-thickness ratio squared, yielding $Q$ factors as high as $10^8$. Various features of these resonators spark the imagination. For example, a beam 400 $\upmu$m wide and 75 nm thick was found to support torsional modes with $Q$-frequency products exceeding $6\times 10^{12}$ Hz (a thermal decoherence time exceeding one mechanical period) and its rotation was resolved using an optical lever with an imprecision 20 dB below that at the Standard Quantum Limit, signaling the potential for a new branch of torsional quantum optomechanics \cite{aspelmeyer2014cavity}.  We also found that strained nanobeams can be mass-loaded without affecting their torsional $Q$, yielding chip-scale torsion pendula with $Q$-$m$ factors as high as 0.1 kg (for the 0.1 mg device in Fig. 4) and damping rates as low as 10 $\upmu$Hz.  We showed how such a device might be used to create a chip-scale micro-$g$ gravimeter (competing in size and simplicity with recent MEMS proposals that employ much larger masses and more complex spring geometries 
\cite{Middlemiss2016, Tang2019, Mustafazade2020}).  We also note that its torque sensitivity, $10^{-19}\;\t{N m}/\sqrt{\t{Hz}}$ \cite{SI}, is on par with the best that has been achieved with milligram test masses using optical levitation \cite{Komori2020}, and as such might be applicable to recent proposals for probing the interface between gravitational and quantum physics \cite{bassi2003dynamical,carney2021mechanical}. 
Finally, we re-emphasize that our measurements indicate that torsion modes of a strained nanobeam are naturally soft-clamped, enabling near ideal dissipation dilution factors---for both the fundamental and higher order modes---without the need for sophisticated mode-shape engineering techniques such as phononic crystal \cite{tsaturyan2017ultracoherent,ghadimi2018elastic} and fractal patterning \cite{beccari2021hierarchical}.  This finding suggests that the landscape for nanoscale dissipation dilution---torsional and otherwise---remains largely unexplored. 

\vspace{1mm}

This work is supported by the NSF Convergence Accelerator Program under Grant 2134830.  DJW acknowledges additional support from NSF Grant ECCS-1945832. CMP acknowledges support from an Amherst College Fellowship and a Grant in Aid of Research from Sigma Xi. ARA and CAC both acknowledge support from Friends of Tucson Optics Endowed Scholarships. Finally, the reactive ion etcher used for this study was funded by an NSF MRI grant, ECCS-1725571.

\vspace{-1mm}
\bibliographystyle{apsrev4-1}
\bibliography{ref}
\end{document}



\title{Supporting Information for\\``Nanoscale Torsional Dissipation Dilution for Quantum Experiments and Precision Measurement''}

\author{J. R. Pratt}
\affiliation{National Institute of Standards and Technology, 100 Bureau Drive, Gaithersburg, MD 20899}

\author{A. R. Agrawal}
\author{C. A. Condos}
\author{C. M. Pluchar}
\affiliation{Wyant College of Optical Sciences, University of Arizona, Tucson, AZ 85721, USA}

\author{S. Schlamminger}%
\affiliation{National Institute of Standards and Technology, 100 Bureau Drive, Gaithersburg, MD 20899}

\author{D. J. Wilson}
\affiliation{Wyant College of Optical Sciences, University of Arizona, Tucson, AZ 85721, USA}



\maketitle





\section{Dissipation dilution of a torsion ribbon}

In this section we derive the central result of the main text: tensioning a ribbon increases its torsional stiffness and, proportionately, the quality factor of its torsion modes:
\begin{equation}\label{eq:dissipationdilution}
\frac{Q}{Q_0}=1+\frac{k_\sigma}{k_E}\approx \frac{\sigma}{2E}\left(\frac{w}{h}\right)^2
\end{equation}

We derive Eq. S1 in two ways. First we provide a heuristic derivation based on on the bifilar theory of Buckley \cite{Buckley1914}, which reveals that the twisting of a tensioned ribbon does not change its length---and therefore magnitude of the tension---to first order.  This geometric nonlinearity is what accounts for the losslessnes of the torsion constant due to tension.  Moving beyond the lump mass model, we provide a full continuum mechanics model, following the generalized dissipation dilution theory of Federov, \emph{et. al.} \cite{Federov2019_Generalized}, that confirms the heuristic result while accounting for the mode-shape dependence of Eq. S1.  Comparing to transverse flexural modes of the ribbon, we find that torsional modes are naturally ``soft-clamped." 

\subsection{Lumped mass model: The bifilar effect}

Quinn, et. al. \cite{Quinn1997}  used a ribbon as the torsion fiber in a balance apparatus designed for measurement of the universal gravitation constant. The $Q$ of this torsion pendulum was enhanced via the bifilar effect described by Buckley \cite{Buckley1914}, and this macroscopic example of dissipation dilution was the inspiration for our investigation of nanomechanical torsion. In short 
we reasoned that a bifilar effect might also offer a path to dissipation dilution in the torsion of nanoribbons. 

\begin{figure}[b!]
\vspace{-4mm}
        \includegraphics[width=1\columnwidth]{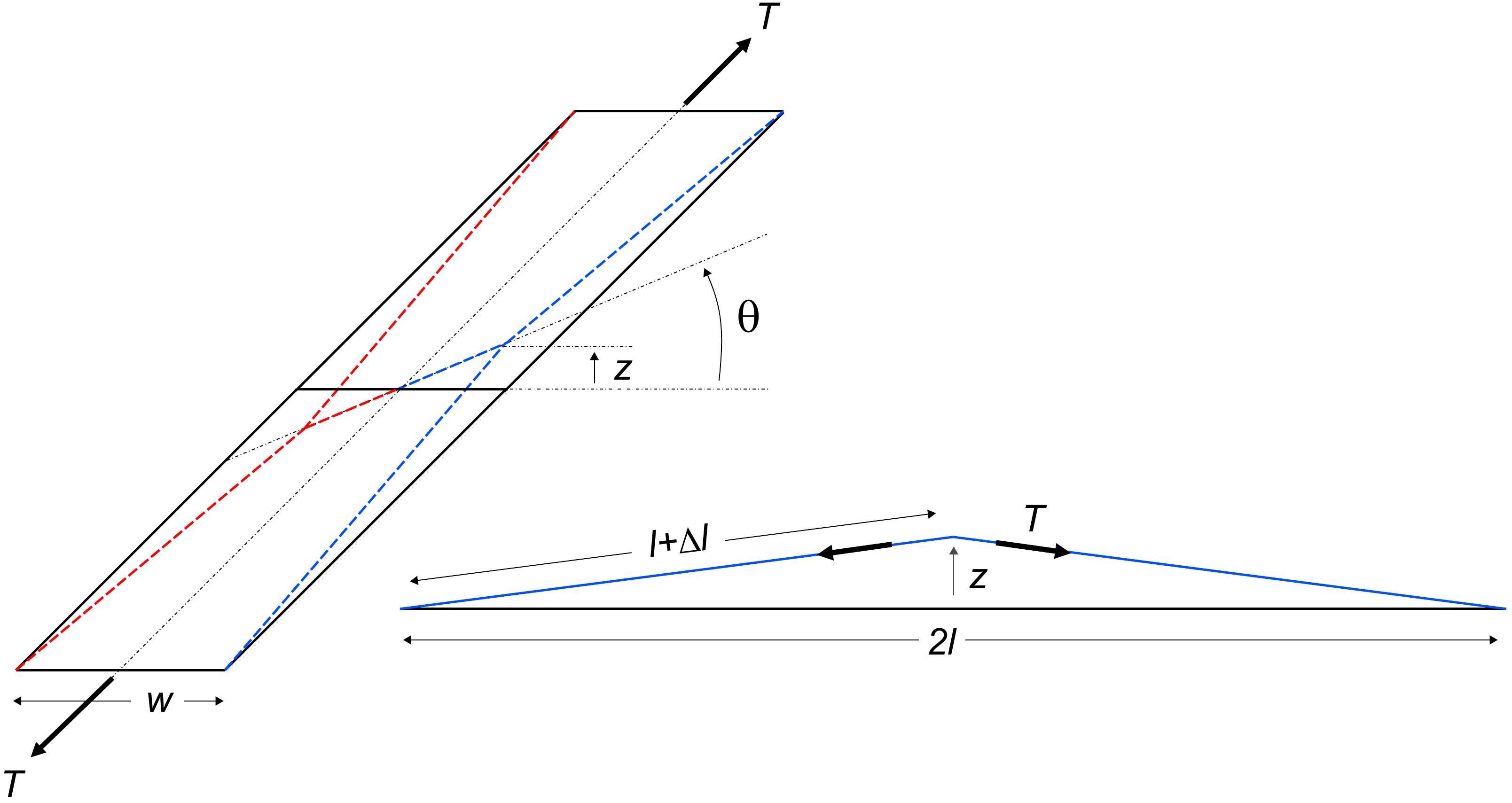}
   \caption{Twisting of a ribbon. The drawing on the left shows a ribbon in equilibrium. The blue dashed lines indicate the excursion of the ribbon's edge if the middle of the ribbon twists. The triangle on the lower right shows the geometry of the edge. }
    \label{fig2}
\vspace{-1mm}
\end{figure}

To retrace our reasoning, consider that the restoring torque $\kappa$ for the ribbon of Fig. \ref{fig2}a is known from torsion balance experiments (e.g. \cite{Quinn1997}) to consist of two components
\begin{equation}\label{eq:torsionstiffness}
  \kappa =\kappa_E + \kappa_{\sigma}.
\end{equation} 
The first component, attributable to Saint-Venant \cite{SaintVentant1856memoire,timoshenko1951theory},
\begin{equation}\label{eq:torsionstiffnessE}
    \kappa_E = \frac{E h^3 w}{6 l}\theta,
\end{equation}
is due to shear deformation the ribbon and is a source of loss. The second component, attributable to Buckley \cite{Buckley1914},
\begin{equation}\label{eq:torsionstiffnessT}
    \kappa_\sigma = \frac{\sigma h w^3}{12 l}\theta,
\end{equation}
is due to the tensile load $T = \sigma h w$, and is lossless.

The lossless nature of the tensile component of the restoring torque is perhaps most easily appreciated in terms of a bifilar suspension, such as a child's swing. When we twist a swing about its longitudinal axis, we tend to lift the mass of its passenger (a child, perhaps) in a screw-like fashion, doing work against gravity. When we release the swing, it unwinds, gathering speed, spinning the passenger until completely unwound,  continuing until it lifts the passenger to the original height, reverses, and starts the oscillation again. The behavior is that of a simple pendulum in the limit of small angles, and in the absence of friction would continue forever.

A key assumption in our description of the child's swing is that the suspensions are inelastic. As such, the potential energy of the child's motion is stored entirely in the conservative gravitational field. Buckley's insight was that this assumption could be extended to a elastic ribbon suspension, by partitioning the ribbon into a set of infinitesimally thin pairs of strings, each behaving like a bifilar suspension. Constraining each suspension to lift the mass by the same amount entails a shear deformation of the ribbon, and thereby an extra elastic restoring torque $\kappa_E$.  For a thin, wide ribbon, however, the dominant restoring torque $\kappa_\sigma$ is due to gravity, or equivalenty, the tensile pre-stress $\sigma$ produced by loading the elastic ribbon. 

As an extension of Buckley's treatment, we now consider a torsion balance that uses fully constrained ribbons (clamped on each end) whose tensile load is not due to the weight of the torsion paddle, but rather due to thin film residual stress. Our aim is to show that the restoring torque due to this stess is lossless.  As such, we must consider if the added constraint leads to stretching of the ribbon, a deformation with attendant loss beyond the aforementioned shear term. 
Heuristically, we follow Buckley, and consider each ribbon to be composed of parallel infinitesimal strings of length $l$ and infinitesimal width $dr$, located a distance $r$ from the center line. Each string is fastened at one end to the fixed frame and at the other end to the torsion paddle, which is free to rotate through angle $\theta$.

 The imagined strings taken as a whole represent a very high aspect ratio ($l/h > 10^4$) membrane of thickness $h$ and width $w$ loaded in tension from residual stress $\sigma$.  The stress is equivalent to a tensile force $T=\sigma A$, where $A=hw$. The tensile force per breadth of ribbon $T_0=T/w$ imposes a tensile load $T_0 dr$ along the length of each infinitesimal string. 
 
 To analyze the torsional stiffness of the ribbon, we consider rotating the paddle through an angle $\theta$. As illustrated in Fig. S1, each infinitesimal string displaces vertically from equilibrium a distance $\Delta z \approx \theta r$ at the paddle end.  It also elongates by an amount $\Delta l \approx \Delta z^2/(2l)$, yielding a (longitudinal) strain
\begin{equation}\label{eq:geometricnonlinearity}
  \epsilon = \frac{\Delta l}{l} \approx \frac{1}{2}\frac{r^2\theta^2}{l^2}
\end{equation} 

The quadratic dependence of strain on rotation angle is responsible for the apparent losslessness of the bifilar effect, since it implies that the restoring torque due to tension contains negligible contribution from deformation. To see this, note that the longitudinal stiffness per breadth of the ribbon is
\begin{equation}
 k=\frac{EA}{wl} 
\end{equation}
where $E$ is the modulus of elasticity. The change in tension of each string due to elongation is then 
\begin{equation}
 \Delta T_0=k dr \Delta L =\frac{EA dr}{w}\epsilon.
\end{equation}
and the vertical restoring force applied by each string to the paddle (containing contributions from both tension $T_0$ and elongation $\Delta T_0$) is
\begin{subequations}\begin{align}
F_z &= (T_0 + \Delta T_0) dr \sin\alpha\\
 & \approx (T_0 + \frac{EA }{w}\epsilon) \frac{r\theta dr}{l(1+\epsilon)}\\
 & \approx T_0\frac{r\theta}{l}\mathrm{d}r  + \frac{1}{2}\bigg(\frac{EA}{w}-T_0\bigg)\frac{r^3\theta^3}{l^3}\mathrm{d}r,
\end{align}\end{subequations} 
where $\alpha\approx r\theta/l $ is the angle of each string from horizontal.  

To compute the total restoring torque on the paddle, we sum the contribution $\kappa_0 \approx F_z r$ from each string, remembering they occur in pairs, one on each side of the neutral line:
\begin{equation}
 \kappa_\sigma \approx  \int_0^{w/2}  2 F_z r dr
 = \frac{T_0}{12} \frac{w^3\theta}{l} +
 \frac{1}{160}\bigg(\frac{EA}{w}-T_0\bigg)\frac{w^5\theta^3}{l^3}
\end{equation}
Substituting $\sigma h$ for $T_0$, the torsional stiffness is
 \begin{equation}
 k_\sigma = \frac{d\kappa_\sigma}{d\theta}\approx \frac{\sigma}{12} \frac{hw^3}{l} +
 \frac{1}{160}\bigg(E-\sigma\bigg)\frac{hw^5\theta^2}{l^3}
 \end{equation}
and we see we have recovered the result due to Buckley, with an additional term due to the linear longitudinal strain that is nonlinear in the displacement $\theta$. The term that is nonlinear in $\theta$ is negligible in the usual small angle limit, but likely important for large deflections, which we hope to investigate.

\subsection{Continuum mechanics model}

We now derive Eq. \ref{eq:dissipationdilution} using a continuum mechanics
model recently developed for nanomechanical resonators \cite{Federov2019_Generalized}, and compare the dilution factor for 
flexural and torsional modes of a rectangular beam. We find that the $Q$ factor of torsional modes should
scale as the beam's aspect ratio squared, mirroring the behavior of ``soft-clamped" flexural modes.

Following \cite{Federov2019_Generalized}, consider an elastic solid with modulus $E = E_0(1+iQ_0^{-1})$ subject to a static stress field $\overline{\sigma}_{ij}$.  Vibrations of the solid are described by the displacement field
\begin{equation}
U_i(\vec{r},t)=\overline{U}_i(\vec{r})+u_i(\vec{r},t)
\end{equation} 
where $\overline{U}_i$ is the static deformation produced by $\overline{\sigma}_{ij}$ and \begin{equation}
u_i(\vec{r},t)=\sum_n \phi_i^{(n)}(\vec{r})A_ne^{i\omega_n t}
\end{equation}
is the time-dependent vibration decomposed into normal modes with frequency $\omega_n$ and mode shape $\phi_i^{(n)}$.

The relationship between the stress and displacement fields is given by the strain tensor (here in Cartesian coordinates)
\begin{equation}
\epsilon_{ij}=\frac{1}{2}\left(\partial_i u_j + \partial_j u_i + \partial_i u_k \partial_j u_k \right)
\end{equation}
and the constitutive relation (Hooke's law)
\begin{equation}
\sigma_{ij} = E\epsilon_{ij}
\end{equation}

We wish to determine the quality factor $Q_n$ of mode $n$, defined as the ratio of the  time-averaged energy stored in the mode $\langle W^{(n)} \rangle$ to the energy dissipated per cycle $\langle W_\t{diss}^{(n)} \rangle$
\begin{equation}
Q_n \equiv \frac{\langle W^{(n)} \rangle}{\langle W_\t{diss}^{(n)} \rangle}
\end{equation}
Towards this end, identify the strain energy of the solid as
\begin{equation}\label{eq:strainenergy}
W(t)=\frac{1}{2}\int \sigma_{ij}(t)\epsilon_{ij}(t)dV
\end{equation}
and the power dissipated as
\begin{subequations}\begin{align}
P_\t{diss}(t)&=\int \sigma_{ij}(t)\dot{\epsilon}_{ij}(t)dV\\
&=\int \sigma_{ij}(t)\dot{\Delta\epsilon}_{ij}(t)dV
\end{align}\label{eq:dissipation}\end{subequations}
where 
\begin{equation}
\Delta\epsilon_{ij}(t) = \epsilon_{ij}(t) - \overline{\epsilon}_{ij}
\end{equation}
is the time-dependent part of the strain tensor.


Some manipulation of Eqs. \ref{eq:strainenergy}-\ref{eq:dissipation} gives
\begin{equation}
\langle W^{(n)} \rangle = E\int\left(\langle\overline{\epsilon}_{ij}\Delta\epsilon_{ij}^{(n)}\rangle+\frac{1}{2}\langle\Delta\epsilon_{ij}^{(n)}\Delta\epsilon_{ij}^{(n)}\rangle\right) dV
\end{equation}
and 
\begin{equation}
\langle W_\t{diss}^{(n)} \rangle=Q_0^{-1}E\int \langle\Delta\epsilon_{ij}^{(n)} \Delta\epsilon_{ij}^{(n)}\rangle dV,
\end{equation}
yielding
\begin{equation}\label{eq:DQ}
\frac{Q^{(n)}}{Q_0} = 1+\frac{\int \overline{\epsilon}_{ij}\langle\Delta\epsilon^{(n)}_{ij}\rangle dV}{\tfrac{1}{2}\int \langle\Delta\epsilon^{(n)}_{ij}\Delta\epsilon^{(n)}_{ij}\rangle dV}\equiv D^{(n)}_Q
\end{equation}

Eq. \ref{eq:DQ} reveals that static strain $\overline{\epsilon}_{ij}$ gives rise to dissipation dilution ($D_Q>1$) in the presence of a geometric strain nonlinearity, $\langle\Delta\epsilon_{ij}\rangle>0$.  In more physical terms, we can identify
\begin{equation}
\langle W_\sigma^{(n)}\rangle \equiv \int\overline{\sigma}_{ij}\langle \Delta\epsilon_{ij}(t)\rangle dV \equiv \frac{1}{2}k_\sigma^{(n)}A_n^2
\end{equation}
as an effective lossless potential due to static stress and 
\begin{equation}
\langle W_E^{(n)}\rangle \equiv \frac{1}{2}E\int \langle\Delta\epsilon^{(n)}_{ij}(t)\Delta\epsilon^{(n)}_{ij}(t)\rangle dV \equiv \frac{1}{2}k_E^{(n)}A_n^2
\end{equation}
as an effective lossy potential due to elastic deformation, with associated spring constants $k_\sigma$ and $k_E$, respectively, and
\begin{equation}\label{eq:generalizedDD}
D_Q^{(n)}=1+\frac{\langle W_\sigma^{(n)}\rangle}{\langle W_E^{(n)}\rangle}=1+\frac{k_\sigma^{(n)}}{k_E^{(n)}}
\end{equation}

We now apply the above formalsim to modes of a tensile-strained rectangular beam.
\subsubsection{Flexural modes of a rectangular beam}

As a well-studied base case, consider flexural modes of a doubly-clamped beam of thickness $h$, width $w$, and length $L\gg{h,w}$ oriented along the $y$, $x$, and $z$ axis, respectively. The beam is subject to a static tensile stress $\overline{\sigma}_{zz} = \sigma$, yielding string-like flexural vibrations along principle axis $y$ of the form
\begin{equation}\label{eq:modeshape_string}
\phi_y^{(n)}(z) = \sin(k_nz)+\phi_{y,\t{clamp}}^{(n)}(z)
\end{equation}
where $k_n = \pi n/L$ and $\phi_{y,\t{clamp}}(z)$ is a correction to the ideal string modeshape which ensures satifaction of the boundary conditions $\phi_y(0)=\phi_y(L)=\partial_z\phi_y(0)=\partial_z\phi_y(L) = 0$. 

To compute $D_Q^{(n)}$, first recognize that only the axial component of the strain tensor is relevant 
\begin{equation}
\Delta\epsilon_{zz} \approx \frac{\partial u_z}{\partial z}+\frac{1}{2}\left(\frac{\partial u_y}{\partial z}\right)^2.
\end{equation}
It is tempting to ignore the leading term on the grounds that the vibration is transverse to $\hat{z}$; however, the finite thickness of the beam introduces an axial strain due to curvature:
\begin{equation}
\frac{\partial u_z}{\partial z}\approx-\frac{\partial^2 u_y}{\partial z^2}y
\end{equation}
This term vanishes from from $\langle W_\sigma \rangle $ (because $\langle u \rangle = 0$) and dominates for $\langle W_E \rangle$, yielding
\begin{subequations}\begin{align}
\langle W_\sigma \rangle &= \frac{\sigma A}{2}\int \langle\left(\frac{\partial u_y}{\partial z}\right)^2\rangle dz\\
\langle W_E \rangle &= \frac{EI}{2} \int \langle\left(\frac{\partial^2 u_y}{\partial z^2}\right)^2\rangle dz
\end{align}\end{subequations}
where $A=wh$ and $I = wh^3/12$ are the area and area moment of inertia of the beam, respectively. 

Evaluating for modeshape $\phi^{(n)}_y(z)$ gives 
\begin{equation}
k_\sigma^{(n)}= \sigma A \frac{ k_n^2L}{2} = \frac{2\sigma hw}{L}\left(\frac{\pi n}{2}\right)^2 
\end{equation}
and 
\begin{equation}
k_E^{(n)}  = k_{E,\t{free}}^{(n)} + k_{E,\t{clamp}}^{(n)}
\end{equation}
where
\begin{equation}
k_{E,\t{free}}^{(n)} = EI\frac{k_n^4L}{2} = \frac{2Ewh^3}{3L^3}\left(\frac{\pi n}{2}\right)^4
\end{equation}
is the stiffness due to curvature at antinodes (the stiffness of a free-free beam) and
\begin{equation}
k_{E,\t{clamp}}^{(n)} = EI\int\left(\tfrac{\partial^2\phi_\t{clamp}^{(n)}}{\partial z^2}\right)^2dz
\end{equation}
is the stiffness due to curvature at the clamps. An approximate $\phi_\t{clamp}^{(n)}$ is given by smoothly transitioning to the mode shape of a cantilever with length $L_c = \sqrt{2EI/\sigma A}$ \cite{schmid2011damping}:
\begin{equation}
\phi^{(n)}_\t{clamp}(z) \approx -k_n L_c\left(\frac{z}{L_c}-\frac{z^2}{L_c^2}+\frac{z^3}{3L_c^3}\right),
\end{equation}
yielding
\begin{subequations}\begin{align}
k_{E,\t{clamp}}^{(n)} &\approx 2EI\int_0^{L_c}\left(\frac{\partial^2\phi_\t{clamp}^{(n)}}{\partial z^2}\right)^2dz\\
&=\frac{8EIk_n^2}{3L_c} = \sqrt{\frac{EI\sigma A}{9/32}}k_n^2\approx\frac{2wh^2}{L^2}\sqrt{E\sigma}\left(\tfrac{\pi n}{2}\right)^2
\end{align}\end{subequations}
Collecting terms yields a dissipation dilution factor of \cite{Federov2019_Generalized,villanueva2014evidence,schmid2011damping}
\begin{subequations}\begin{align}
D_Q^{(n)}&=1+\frac{k_\sigma^{(n)}}{k_{E,\t{free}}^{(n)}+k_{E,\t{clamp}}^{(n)}}\\
&\approx 1+\left(\frac{E}{\sigma}\left(\frac{h}{L}\right)^2\frac{\pi^2 n^2}{12}+1.09\sqrt{\frac{E}{\sigma}}\frac{h}{L}\right)^{-1}\label{eq:DQtorsion}
\end{align}\end{subequations}
which is bound by the ``soft-clamping" limit 
\begin{equation}
D_{Q,\t{SC}}^{(n)} =1+\frac{k_\sigma^{(n)}}{k_{E,\t{free}}^{(n)}}<\frac{\sigma}{E }\left(\frac{L}{h}\right)^2\frac{12}{\pi^2 n^2}.
\end{equation}
\subsubsection{Torsional modes of a rectangular beam}
We now consider torsion modes of a tensile-stressed rectangular beam, \mbox{which we assume take the soft-clamped form \cite{timoshenko1951theory}}
\begin{equation}\label{eq:modeshape_torsion_cylindrical}
\phi_\theta^{(n)}(z) = \sin(k_nz)
\end{equation}
where $\theta$ denotes the angle of rotation about the beam axis $z$.

To compute $D_Q^{(n)}$, we use the same procedure as before; however, the contribution of shear stresses must be considered.  Following the approach of Saint-Venant \cite{SaintVentant1856memoire} with $\phi_\theta\ll 1$ and ``warping function" $W(x,y)$, assume that the mode shape can be expressed in Cartesian coordinates as  \cite{chopin2019extreme,sapountzakis2013bars, love2013treatise}
\begin{subequations}\label{eq:modeshape_torsion_cartesian}\begin{align}
\phi_x^{(n)}(x,y,z) & = -y\phi_\theta^{(n)}(z)\\
\phi_y^{(n)}(x,y,z) & = x\phi_\theta^{(n)}(z)\\
\phi_z^{(n)}(x,y,z) & = W(x,y)(\phi_\theta^{(n)}(z))'_z
\end{align}\end{subequations}
with the associated strains
\begin{subequations}\label{eq:axialstrain_torsion}\begin{align}
\Delta\epsilon_{zz} &= W(x,y)\frac{\partial^2u_\theta}{\partial z^2}+\frac{1}{2}(x^2+y^2)\left(\frac{\partial u_\theta}{\partial z}\right)^2\\
\Delta \epsilon_{xz}&=\frac{1}{2}\left(\frac{\partial W}{\partial x}-y\right)\frac{\partial u_\theta}{\partial z}=\Delta \epsilon_{zx}\\
\Delta \epsilon_{yz}&=\frac{1}{2}\left(\frac{\partial W}{\partial y}+x\right)\frac{\partial u_\theta}{\partial z}=\Delta \epsilon_{zy}.
\end{align}\end{subequations}

The warping function of a thin beam ($h\ll w$) with a constant twist rate $(u_\theta)'_z = 0$ is known to be $W(x,y)\approx -xy$, and we will use it here as an approximation to obtain
\begin{subequations}\label{eq:torsionstrain_thinbeam}\begin{align}
\Delta\epsilon_{zz} &\approx -xy\frac{\partial^2u_\theta}{\partial z^2}+\frac{1}{2}x^2\left(\frac{\partial u_\theta}{\partial z}\right)^2\\
\Delta \epsilon_{xz}&\approx -y\frac{\partial u_\theta}{\partial z}\\
\Delta \epsilon_{yz}&\approx 0.
\end{align}\end{subequations}
(Note we have dropped the term $y^2 (u_\theta)''_z$ from $\Delta\epsilon_{zz}$ as it contributes negligibly to the total strain energy for $h\ll w$). 

By inspection, $\Delta\epsilon_{zz}$ is identical to that for a flexural mode; however, rotation gives rise to an additional shear strain $\epsilon_{xz}=-2y(u_\theta)'_z$. Only the nonlinear term in $\Delta\epsilon_{zz}$ contributes to $\langle W_\sigma\rangle$, whereas the shear term dominates $\langle W_E\rangle$, yielding
\begin{subequations}\begin{align}
	\langle W_\sigma \rangle &= \frac{\sigma}{2}\frac{hw^3}{12}\int \langle\left(\frac{\partial u_\theta}{\partial z}\right)^2\rangle dz\\
	\langle W_E \rangle &= \frac{E}{2}\frac{h^3w}{12}\int\langle 2\left(\frac{\partial u_\theta}{\partial z}\right)^2 + \frac{w^2}{12}\left(\frac{\partial^2 u_\theta}{\partial z^2}\right)^2\rangle dz
\end{align}\end{subequations}

Evaluating for modeshape $\phi_y^{(n)}(x,z)$ gives
\begin{equation}
k_\sigma^{(n)} = \sigma \frac{hw^3}{12} \frac{k_n^2L}{2} =  \frac{\sigma hw^3}{6L}\left(\frac{\pi n }{2}\right)^2 
\end{equation}
and 
\begin{equation}
k_E^{(n)}  = k_{E,\t{free-shear}}^{(n)} + k_{E,\t{free-bend}}^{(n)}
\end{equation}
where
\begin{equation}
	k_{E,\t{free-shear}}^{(n)}  = E \frac{h^3w}{6} \frac{k_n^2L}{2} =  \frac{E}{3}\frac{h^3w}{L} \left(\frac{\pi n }{2}\right)^2
\end{equation}
is the stiffness due to shear deformation and
\begin{equation}
k_{E,\t{free-bend}}^{(n)}  = E \frac{h^3w^3}{(12)^2} \frac{k_n^4L}{2} =  \frac{3E}{2} \left(\frac{hw}{3L}\right)^3 \left(\frac{\pi n}{2}\right)^4 
	\end{equation}
is the stiffness due to distributed curvature.  

Collecting terms yields a dissipation dilution factor of 
\begin{subequations}\begin{align}
	D_Q^{(n)}&=1+\frac{k_\sigma^{(n)}}{k_{E,\t{free-shear}}^{(n)}+k_{E,\t{free-bend}}^{(n)}}\\
	&=1+\frac{\sigma}{2E}\left(\frac{w}{h}\right)^2\left(1+\frac{1}{6}\left(\frac{\pi n w}{2L}\right)^2\right)^{-1}.
	\end{align}\end{subequations}
which exhibits ``soft-clamped" scaling for $w\lesssim(L/n)$.

\vspace{-3mm}
\section{Miscellaneous properties of torsion modes}
\vspace{-3mm}

In this section we highlight various scalar properties of the torsion modes relevant to the main text  (e.g. effective mass), building on the preceding continuum mechanics model.

\subsection{Resonance frequency and moment of inertia}

Torsion mode frequencies $\omega_n$ can be predicted from the mode stiffness $k_n$ by the formula
\begin{equation}
k_n = k_\sigma^{(n)} + k_E^{(n)} = I_n\omega_n^2
\end{equation}
where 
\begin{equation}
\label{eq:momentOfInteria}
I_n = \frac{\int{(\phi_\theta^{(n)})^2 r_\perp^2 dm}}{(\phi_{\theta,\t{max}}^{(n)})^2} = \frac{\rho (hw^3+h^3w)}{12}\frac{L}{2}\approx \frac{\rho Lhw^3}{24}
\end{equation}
 is the effective moment of inertia of mode $n$.  
 
 For high aspect ratio beams, $\{w,h\}\ll L$, the non-angular resonance frequency of a torsion mode is
 \begin{subequations}\begin{align}
 	\frac{\omega_n}{2\pi}&\approx \frac{1}{2\pi}\sqrt{\frac{k_\sigma^{(n)} + k_{E,\t{free-shear}}^{(n)}}{I}}\\
 	&=\frac{ n}{2L}\sqrt{\frac{\sigma}{\rho}\left(1+\frac{ 4E }{\sigma }\frac{h^2}{w^2}\right)}
 \end{align}\end{subequations}
which is identical to that of a flexural mode in the limit $h\rightarrow 0$.

\subsection{Effective mass}

The effective mass of a torsion mode (defined relative to the point of maximum displacement $\phi_y^\t{max}$)  is given by
\begin{subequations}\begin{align}
m_n &= \frac{\int (\phi_y^{(n)}(x,y,z))^2 dm}{(\phi_{y,\t{max}}^{(n)})^2}\\
&=\rho\frac{\int (x\phi_\theta^{(n)}(z))^2 dxdydz}{(x_\t{max}\phi_{\theta,\t{max}}^{(n)})^2}\\
&=\rho \frac{hw}{3}\int_0^L \sin^2(k_n z) dz\\
\label{eq:effectiveMass}
&=\rho \frac{hwL}{6} = \frac{1}{6}m_\t{phys}
\end{align}\end{subequations}
which is notably 3 times smaller than that of a flexural mode.

Te effective moment of inertia and mass are related by
\begin{equation}
    \label{eq:momentOfInteria2}
    I_n = m_n r_{\perp,\t{max}}^2 = m_n (w/2)^2.
    \end{equation}

\subsection{Zero-point and thermal displacement}

In the limit of high strain ($k_\sigma^{(n)}\gg k_E^{(n)}$), the zero point angular displacement of a torsion mode is given by
\begin{equation}
    \theta^{(n)}_\t{zp} = \sqrt{\frac{\hbar}{2 I_n \omega_n}}
    =\sqrt{\frac{12\hbar }{ h w^3 \pi \sqrt{\rho\sigma}}}
\end{equation}
The resonant zero-point displacement spectral density---Eq. 4 in the main text---is given by
\begin{subequations}\begin{align}
    S_{\theta}^\t{zp,(n)} &= \frac{ 4(\theta^{(n)}_\t{zp})^2}{\gamma_m} = \frac{2\hbar Q_n}{I_n \omega_n^2}\\
    &\le\frac{24L\hbar Q_0[h]}{ h^3 w \pi^2 E}\propto \frac{L}{h^2 w}\label{eq:szp}
\end{align}\end{subequations}
where $\gamma_m = \omega_n/Q_n$ is the mechanical damping rate and to obtain Eq. \ref{eq:szp} we've assumed the dissipation dilution factor in Eq. \ref{eq:DQtorsion} and the surface loss scaling $Q_0[h]\propto h$.

From these expressions we obtain the thermal displacement
\begin{equation}
    \theta^{(n)}_\t{th} = \sqrt{2n_\t{th}^{(n)}}\theta^{(n)}_\t{zp} = \sqrt{\frac{k_B T}{I_n \omega_n^2}}
\end{equation}
and resonant thermal displacement spectral density
\begin{subequations}\begin{align}
    S_{\theta}^\t{th,(n)} &= \frac{ 4(\theta^{(n)}_\t{th})^2}{\gamma_n} = \frac{k_B T Q_n}{I_n \omega_n^3}\\
    &\le\frac{24L^2 k_B T Q_0[h]}{ h^3 w \pi^3 E \sqrt{\sigma/\rho}}\propto \frac{L^2}{h^2 w}.
\end{align}\end{subequations}
where $n_\t{th}^{(n)} = k_B T/\hbar\omega_n$ is the thermal mode occupation.

\subsection{Thermal torque sensitivity}

In the limit of high strain, the thermal torque sensitivity of a torsion mode is given by
\begin{subequations}\begin{align}
	S_{\tau}^{\t{th},(n)}&=4k_B T I_n\gamma_n\\
	&\le\frac{ \pi E k_B T  h^3 w  }{ 3 \sqrt{\sigma/\rho}Q_0[h]}\propto h^2w
\end{align}\end{subequations}
using the surface loss scaling $Q_0[h]\propto h$.  

\subsection{Example: A Si$_3$N$_4$ nanoribbon}
To give an example relevant to the main text, consider a $\{L,w,h\}=\{7\,\t{mm},\,100\,\mu\t{m},\,75\,\t{nm}\}$ Si$_3$N$_4$ beam. Predicted values for the fundamental torsional mode are
\begin{equation}
\frac{\omega_1}{2\pi} = 40\;\t{kHz}\cdot\frac{7\,\t{mm}}{L}\sqrt{\frac{2700\,\tfrac{\t{kg}}{\t{m}^3}}{\rho}\frac{\sigma}{0.85\,\t{GPa}}}
\end{equation}
and 
\begin{equation}
Q_1 =1.4\cdot 10^7\cdot\frac{75\,\t{nm}}{h}\left(\frac{100\,\mu\t{m}}{w}\right)^2\frac{\sigma}{0.85\,\t{GPa}}\frac{250\,\t{GPa}}{E}\frac{Q_0[h]/h}{60/\t{nm}}.
\end{equation}

We measure $\omega_1/2\pi = 40\t{kHz}$ and $Q_1 = 1.6\times10^6$, in good agreement with these predictions. 

Other predicted values are (using $\rho$, $E$, $\sigma$, $Q_0$ as above):
\begin{subequations}\begin{align}
    I_1 & = 5.9\cdot 10^{-20}\,\t{kg\cdot m^2}\cdot\frac{L}{3.5\,\t{mm}}\frac{h}{75\,\t{nm}}\left(\frac{w}{100\,\mu\t{m}}\right)^3\\
	m_1 & = 24\,\t{ng}\cdot\frac{h}{75\,\t{nm}}\frac{w}{100\,\mu\t{m}}\frac{L}{3.5\,\t{mm}}\\
	S_{\theta}^{\t{zp},(1)}&=\left(6.7\cdot10^{-10}\frac{\t{rad}}{\sqrt{\t{Hz}}}\right)^2\cdot\frac{L}{7\,\t{mm}}\left(\frac{75\,\t{nm}}{h}\right)^2\frac{100\,\mu\t{m}}{w}\\
		S_{\theta}^{\t{th},(1)}&=\left(1.2\cdot10^{-5}\frac{\t{rad}}{\sqrt{\t{Hz}}}\right)^2\cdot\left(\frac{L}{7\,\t{mm}}\frac{75\,\t{nm}}{h}\right)^2\frac{100\,\mu\t{m}}{w}\\
		S_{\tau}^{\t{th},(1)}&=\left(4.2\,\frac{\t{zN}\cdot\t{m}}{\sqrt{\t{Hz}}}\right)^2\left(\frac{h}{75\,\t{nm}}\right)^2 \frac{w}{100\,\mu\t{m}}
\end{align}\end{subequations}

\newpage

\section{Fabrication and numerical modeling}

In this section, we provide details on the fabrication and modeling of Si$_3$N$_4$ nanobeams decribed in the main text.  The basic fabrication process flow is shown in Fig. \ref{fig:processflow}.  Also shown is the mask pattern defining the shape of the beam with and without a central paddle for mass-loading.

\subsection{Fabrication}

\subsubsection{Unloaded Si$_3$N$_4$ nanobeams}

Fabrication begins by coating a 1.5-um-thick S1813 positive tone photoresist on a double-sided, 100-nm-thick Si$_3$N$_4$-on-silicon wafer. The resist on one side of the wafer (the front side) is patterned in the shape of diagonal beam using a photolithography system (MLA 150), while the resist on the other side protects the wafer from handling scratches. The pattern is transferred to the Si$_3$N$_4$ thin film using fluorine-based (Ar + SF$_6$) reactive ion dry etching. The remaining resist is then removed and a fresh resist layer is applied to protect the front side of the wafer. The back side is then patterned with square windows while making sure it is aligned with the front side\cite{norte2016mechanical}. After dry etching the back side pattern, the remaining resist is removed and the wafer is cleaned using oxygen plasma to remove any lingering resist residues. A thick layer of resist is then coated on both sides and the wafer is diced into $12\times12$ mm$^2$ chips. The chips are then cleaned using 10 second dip in hydrofluoric acid (HF) followed by a DI water and Isopropanol (IPA) rinse\cite{reinhardt2016ultralow}. Chips are then mounted onto a custom Teflon holder to secure them in a vertical orientation. The assembly is then etched in a potassium hydroxide (KOH) bath at 85 $^\circ$C for 21 hours in order to remove the silicon in the patterned region and subsequently release the beam. The released structure is dried using a gradual dilution process which includes iteratively replacing KOH with DI water followed by a 10 min HF dip and then an IPA and methanol rinse\cite{norte2016mechanical}. Finally, the chips with wider beams are dried in air and the ones with thinner beams are dried using critical point drier.

\begin{figure}[t]
    \includegraphics[width=0.95\columnwidth]{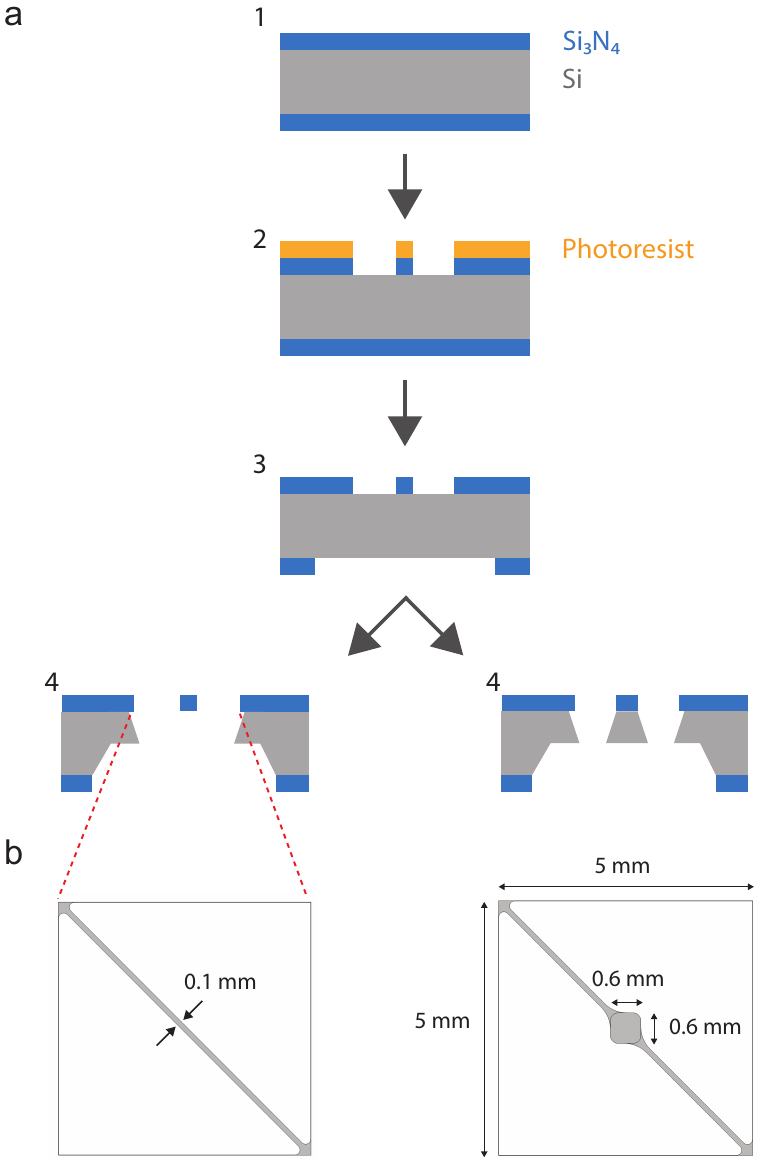}
   \caption{\textbf{Fabrication process flow and mask patterns.} (a) Process flow: (1) double-sided LPCVD Si$_3$N$_4$-on-Si wafer; (2) front side photolithography using S1813 photoresist, followed by Si$_3$N$_4$ dry etch; (3) backside pattern, Si$_3$N$_4$ dry etch, and sample cleaning using O$_2$ plasma and HF dip; (4) KOH wet etch and dry release, resulting in beam with or without mass-loading depending on mask pattern. (b) Mask patterns for unloaded (left) and mass-loaded (right) beams.}
\label{fig:processflow}
\end{figure}

\subsubsection{Mass-loaded Si$_3$N$_4$ nanobeams (torsion microbalances)}

Mass-loaded nanobeams were fabricated using the same procedure as above, except that the the beam pattern includes a $600\times600\;\mu\t{m}^2$ pad in the center. Due to the large pad size and anisotropic etching of Si along <100> crystal plane, in this case there remains an approximately $100$-$\mu\t{m}$-thick Si mass suspended beneath the pad region at the end of the wet etch.

\subsection{Numerical Modeling}
To predict the mode shapes and eigenfrequencies of non-trivial beam geometries---in particular, the torsion modes filleted beams shown in Fig. 2 of the main text---we performed finite element method (FEM) based simulations using the COMSOL 5.4 structure mechanics module. We used the plate physics interface, which allowed us to simulate resonator geometries with large aspect ratios. Unlike the analytical model (Eq. 1 in the main text), which assumes a uniform stress field, we computed the non-uniform stress distribution using a pre-stressed eigenfrequency study. The study is carried out in two steps: the first step is to calculate the von Mises stress distribution due to in-plane stress in the Si$_3$N$_4$ thin film; the second step is to compute eigenfrequencies and their corresponding mode shapes using the stress distribution from step 1\cite{sadeghi2019influence}. All simulations were performed using triangular mesh settings, with dense meshing around the curves and edges. For dissipation dilution models (Sec. \ref{sec:dd}), mesh size was reduced until results changed by less than $1\%$ with successive reduction.

\begin{figure}[h]
\vspace{-4mm}
    \includegraphics[width=0.95\columnwidth]{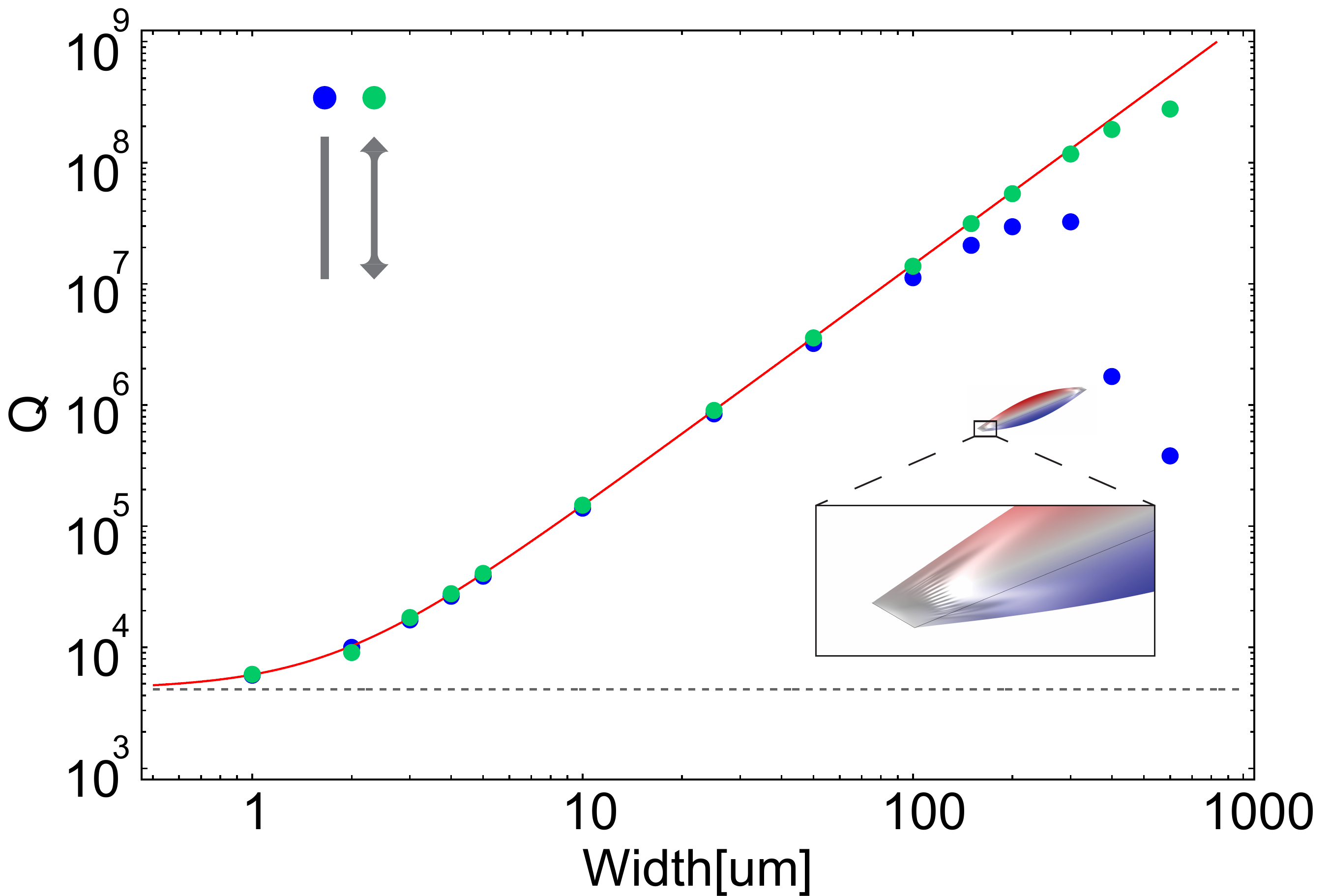}
   \caption{Simulation of quality factors for 75 nm thick nanobeams with (green point) and without (blue point) filleting, assuming surface loss $Q_0[h]=60h/\t{nm}$.  The red curve is the lump mass model given by Eq. 1 in the main text. Inset: ``wrinkling" of the rectangular beam at large widths, coinciding with reduced quality factor.}
\vspace{-2mm}
\label{fig:DD_COMSOL}
\end{figure}

\subsubsection{Dissipation dilution of filleted vs rectangular nanobeams: Buckling instabilities}\label{sec:dd}
The numerical $Q$ model (dashed line) in Fig. 2d of the main text was generated using COMSOL-simulated modeshapes and computing the dissipation dilution factor ($Q/Q_0$) as the ratio of the total kinetic energy stored in the mode (COMSOL function plate.Wk\_tot) to the total elastic strain energy (COMSOL function plate.Ws\_tot).  From these simulations, we predict that the $Q$ factor of torsional modes is highly sensitive to the beam aspect ratio and fillet geometry in the clamps, due to buckling instabilities \cite{kudrolli2018tension,green1937elastic}.  To see this, in Fig. \ref{fig:DD_COMSOL}, we compare the lumped mass model in Fig. 2 (red line) to simulations for filleted beams (green points) and rectangular beams (blue points) of different widths. For the filleted beams, we assume the diagonal geometry of our actual devices (Fig. \ref{fig:processflow}b), which have a fillet radius of $r = 100\;\mu\t{m}$ for beam widths $w<100\;\mu\t{m}$ and $r = w$ for widths $w>100\;\mu\t{m}$.  For rectangular beams, we assume a straight-edged clamp and $r=0$ for all widths.  Both simulations match well to the lump massed model for small widths; however, beyond a critical width, the dissipation dilution factor drops.  Inspection of the modeshope reveals that this coincides with ``wrinkling" of the beam due to buckling instabilities \cite{kudrolli2018tension}.  The use of fillets (ours are inspired by the designs in \cite{sadeghi2019influence}) appears to have the effect of "pulling out" the wrinkles, allowing for higher aspect ratios and concomittently higher $Q$ factors. 

\newpage
\section{Ringdown characterization of nanobeams} 

In this section we provide details on the ringdown measurements shown in Fig. 2 of the main text.

\vspace{-2mm}
\subsection{\label{sec:level1} Experimental Setup}

\begin{figure}[h]
\vspace{-4mm}
    \centering
    \includegraphics[width=0.9\columnwidth]{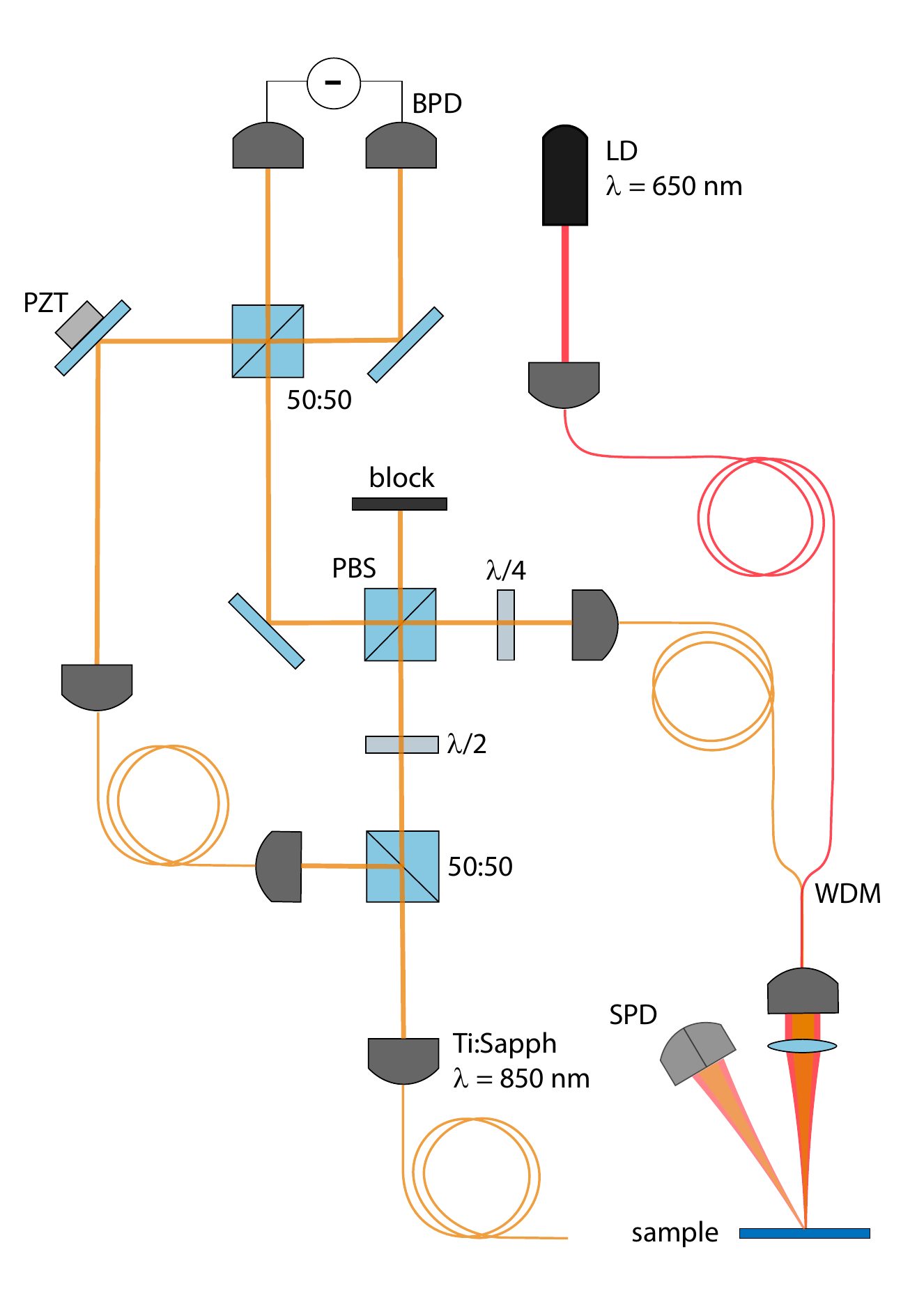}
    \caption{Diagram of the optical setup for ringdown measurements. BPD = balanced photodetector, LD = laser diode, PZT = piezo-electric transducer, PBS = polarizing beam splitter, WDM = wavelength division multiplexer, SPD = split photodiode used in the optical lever configuration.}
    \label{fig:ringdownexperimentalsetup}
\vspace{-2mm}
\end{figure}

For the ringdown measurements shown in Fig. 2 of the main text, Si$_3$N$_4$ nanobeams were housed inside a Kimball Physics 2.75" spherical cube ultra-high-vacuum (UHV) chamber with a typical base pressure of $4\times 10^{-8}$ mbar. The optical setup is shown in Fig. 4 and consists of two readout schemes operating at a nominal wavelength of 850 nm: homodyne interferometry and optical lever.  We alternated between these schemes at different stages of the experiment and obtained the same results.  The balanced photodetector (BPD) used for homodyne readout was a Newport 1807.  The split photodetector (SPD) was a Thorlabs PDQ80.  In addition to the 850 nm probe field (provided by a Titatinum-Sapphire laser, M-Squared SolsTiS), light from a 650 nm diode laser was coupled into the setup via a dichroic fiber beamsplitter (WDM). This laser was used for alignment and radiation pressure actuation.


\subsection{Mode Identification}

\begin{figure}[h!]
    \centering
    \includegraphics[width=0.9\columnwidth]{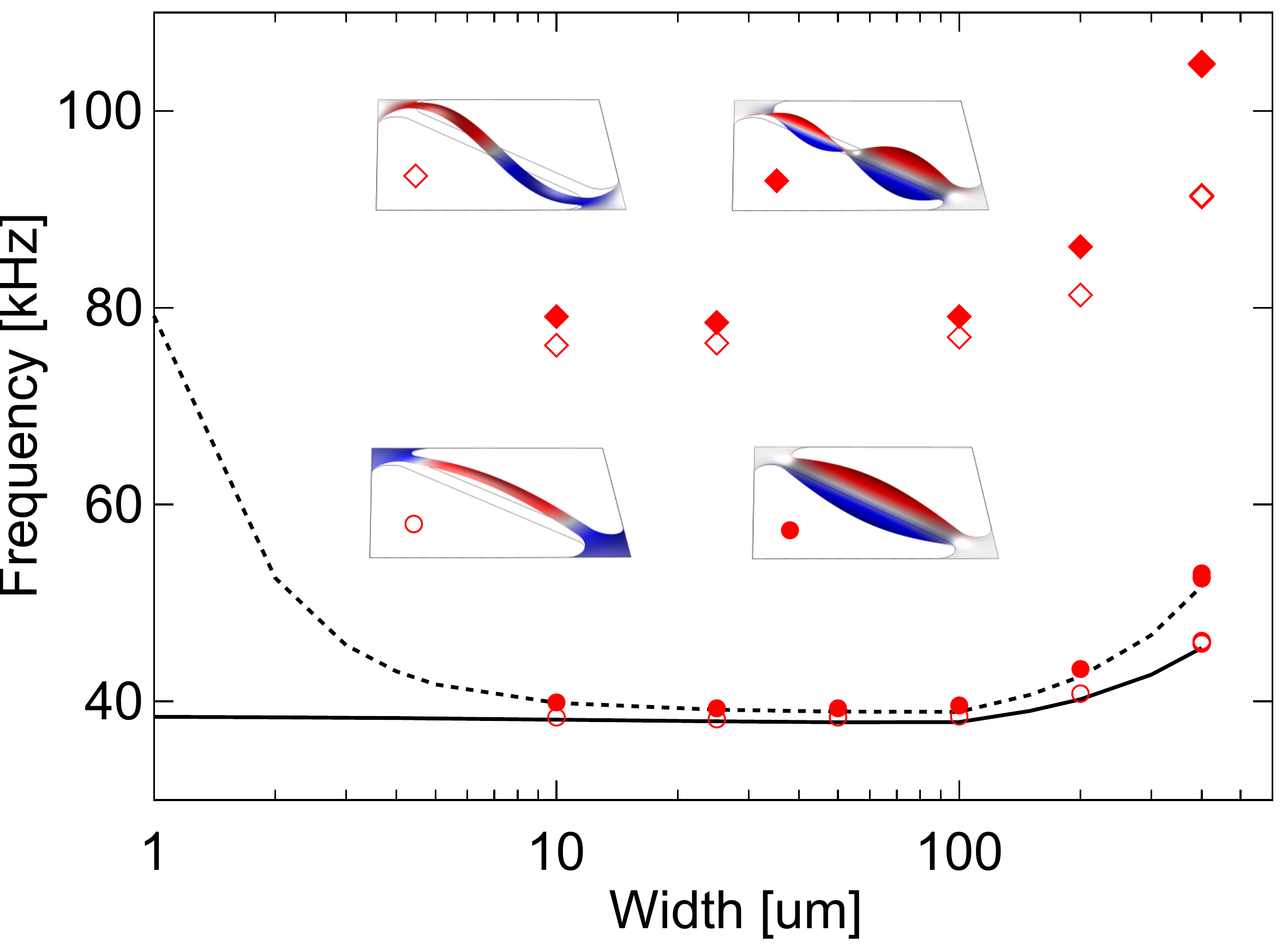}
    \caption{Nanobeam vibrational mode frequency versus beam width. Solid and open markers are measurements for the first order (circle) and second order  (diamond) order torsional and flexural mode, respectively. Solid and dashed lines are COMSOL models for the first order torsional and flexural modes, respectively.}
    \vspace{-2mm}
\label{fig:modefrequencies}
\end{figure}

Before perfoming ringdowns, flexural and torsional modes were identified by comparing resonance peaks in broadband thermal noise spectra to COMSOL silmualations. An example of a thermal noise spectrum is shown in Fig. \ref{fig:higherordermodes}.  For the beams studied in the main text, the fundamental torsional mode frequency was predicted to be $\sim1-10\%$ higher than the fundamental flexural mode frequency.  Measured values were in good agreement, as shown in Fig. \ref{fig:modefrequencies} by plotting measured and simulated mode frequency versus beam width, for both the first and second order flexural and torsional modes.

\subsection{Ringdown Measurements}

Ringdown measurements were performed using the 650 nm diode laser as a radiation pressure actuator\footnote{We note that for torsion modes, the efficiency of the radiation pressure drive was found to be highly sensitive to the alignment of the optical beam.} and either the homodyne or optical lever measurement for readout.  The diode laser was intensity modulated via its current driver using an arbitrary waveform generator (National Instruments PXI 5106) synchronized with a digitizer (National Instruments PXI 5122) recording the photocurrent.  For each measurement, the drive frequency was swept in small increments across mechanical resonance $\omega_n$ while monitoring the photocurrent power spectrum in a window $\Delta\omega\gg \omega_n/Q_n$ centered around $\omega_n$.  When the power rose above a nominal threshold (due to mechanical excitation), the drive beam was shuttered off and the free energy decay of the mode was recorded by tracking the power spectrum as a function of time.

\subsubsection{Investigation of photothermal heating}

\begin{figure}[h]
   \label{fig:photothermalheatingringdown}
    \centering
    \includegraphics[width=0.9\columnwidth]{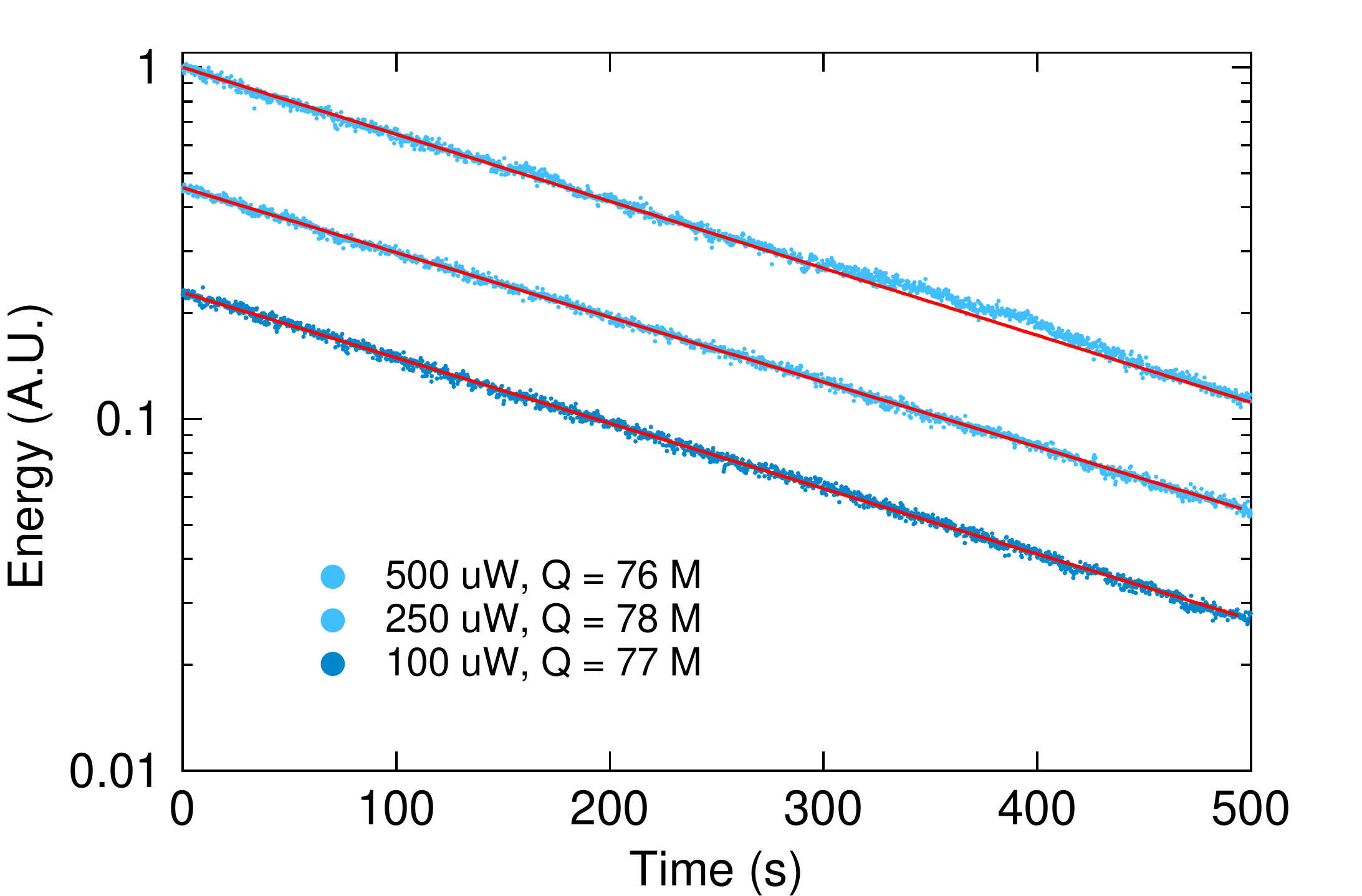}
    \caption{Ringdowns of the fundamental torsion mode of a 400 um nanobeam ($\omega_1 = 2\pi\cdot 52.5$ kHz) with different probe powers.}
\vspace{-3mm}
\end{figure}

For ringdowns of unloaded beams in Fig. 2 of the main text, a typical probe power of 10-50 $\mu$W was used. We observed that at these powers, the ringdown times were not affected by photothermal damping.  For example, in Fig. \ref{fig:photothermalheatingringdown}, we show ringdowns of the fundamental torsional mode of a 400 $\mu$m wide beam for probe powers varying from 100 to 500 $\mu$W, and find that the inferred $Q$ factor ($Q\approx 77\times 10^6$) remains unchanged to within a few percent.

\subsubsection{Comparison of torsional and flexural modes}

\begin{figure}[b]
    \centering
    \includegraphics[width=0.9\columnwidth]{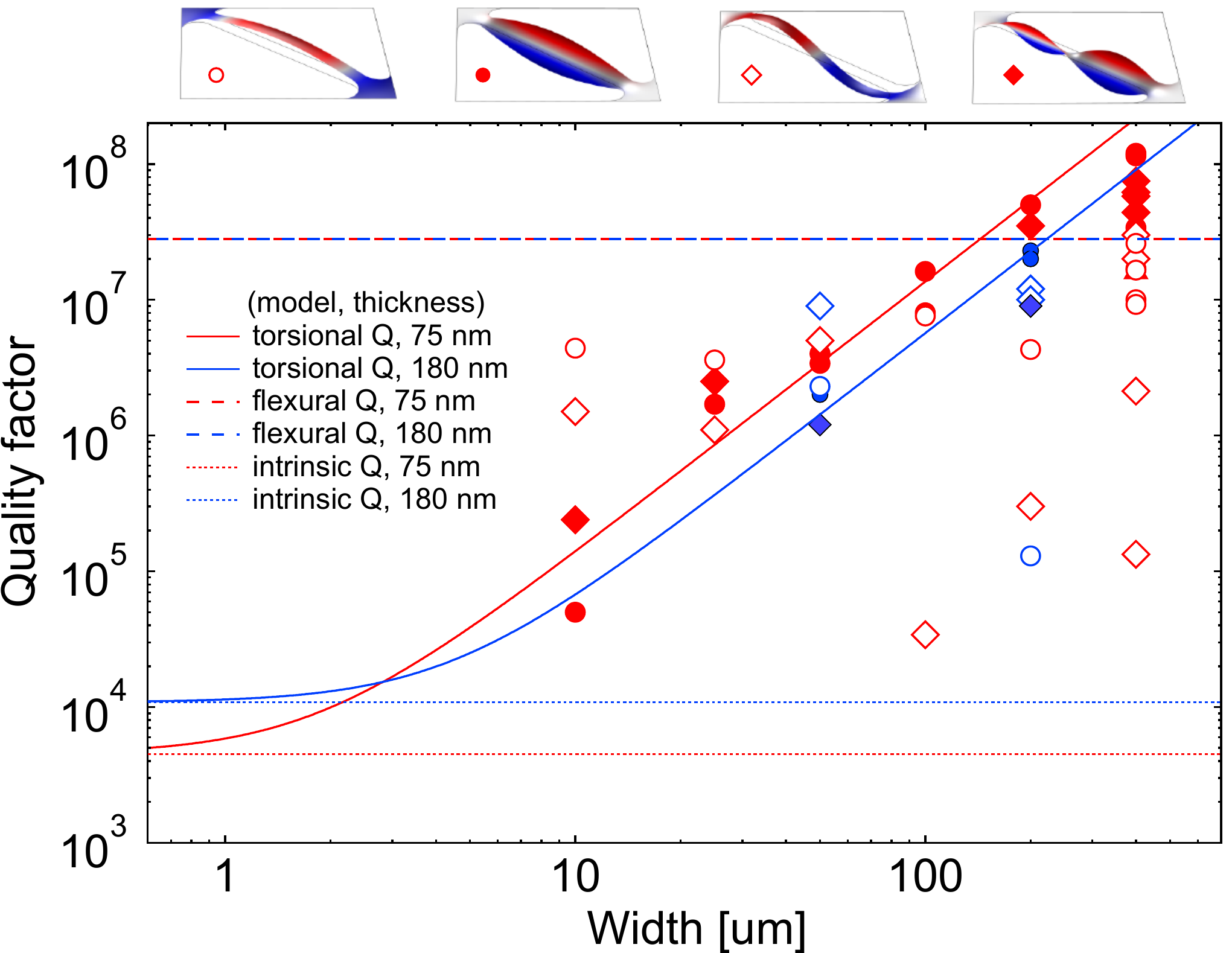}
    \caption{Compilation of quality factors for flexural and torsional modes with different widths and thicknesses.}
          \label{fig:flexuretorsioncomparison}
\end{figure}

As mentioned in the main text, we recorded $Q$ factors for both flexural and torsional modes, and observed qualitatively different scaling versus beam width and thickness, consistent with ``hard-clamping" versus "soft-clamping."  A full compilation of measurements is shown in Fig. \ref{fig:flexuretorsioncomparison}.  Unlike torsional modes (solid markers), flexural modes (open markers) were observed to have roughly constant Q versus width and thickness, as expected for hard-clamping (dashed lines, corresponding to Eq. S36).  The measured Q factors were roughly an order of magnitude lower than predicted, and had significantly larger statistical spread than for torsional modes.  We conjecture that this may be due to larger coupling of flexural modes to low $Q$ modes of the underlying Si frame.

\subsubsection{Investigation of viscous damping}

We investigated the possiblity that our measurments may be limited by gas damping in an attempt to explain the apparent rolloff in nanobeam quality factor evident in Fig. 2d of the main text for widths $\geq$ 400 $\mu m$. Toward this end, we compared the measured quality factors to a model for a nanobeam in the free molecular flow damping regime, a regime for which the mean free path is longer than the largest dimension of the resonator, given by \cite{verbridge2008gasdamping}: 
\begin{equation}
\label{eq:gasdamping}
    Q_{gas} = \frac{\rho t \Omega_m}{4}\sqrt{\frac{\pi}{2}}\sqrt{\frac{RT}{M}}\frac{1}{P}
\end{equation} 

\begin{figure}[ht]
    \centering
    \includegraphics[width=1\columnwidth]{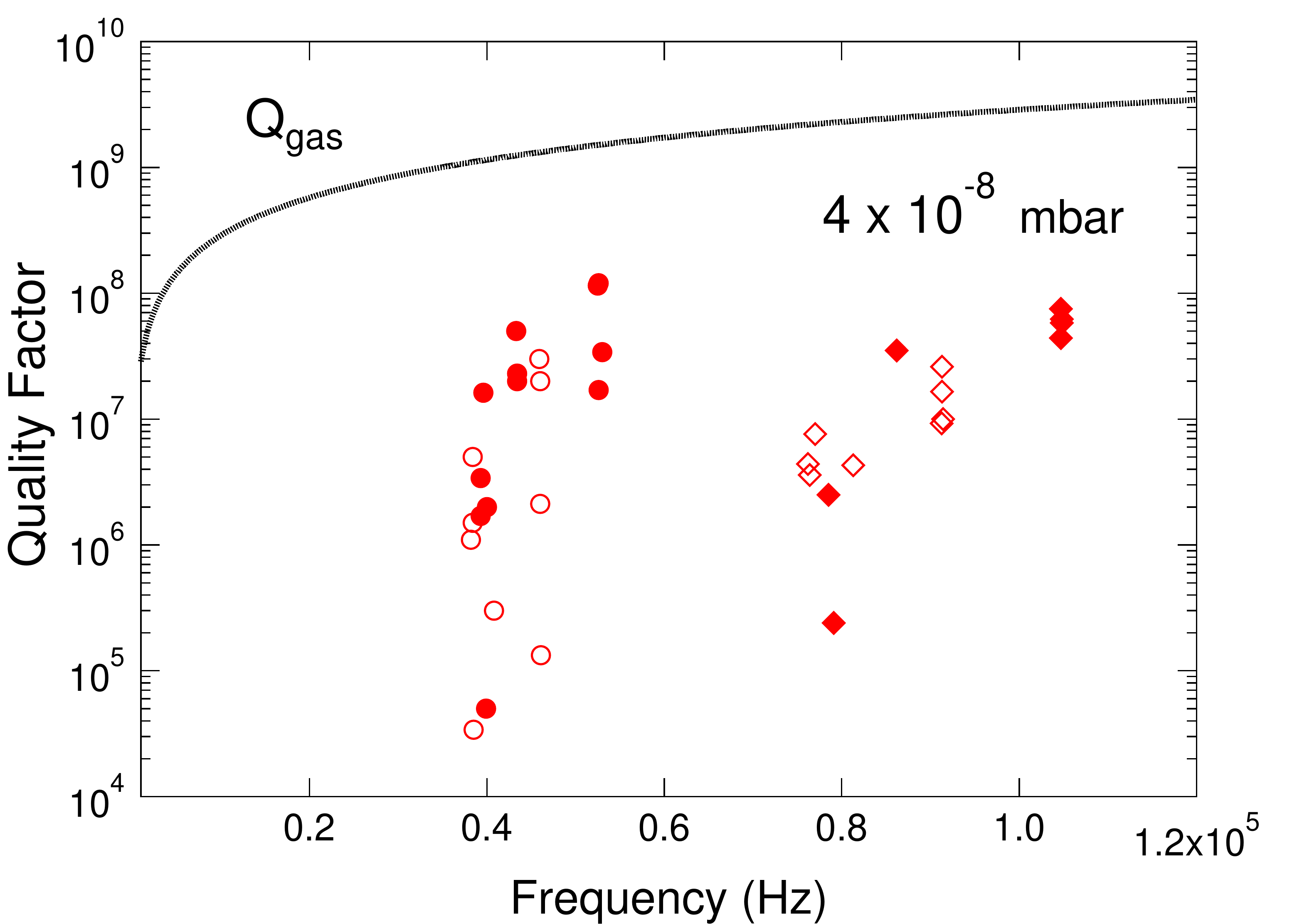}
    \caption{Quality factors of nanobam modes versus resonance frequency, overlaid with a model of the quality factor of a nanobeam limited by gas damping in the free molecular flow regime for $P\approx4\times10^{-8}$ mbar. Markers denote the same modes as in Figs. S5 and S7.}
          \label{fig:gasdamping}
\end{figure}

In this model we assume the density of Si$_3$N$_4$ to be $\rho = $2700 kg/m$^3$, a nanobeam thickness $t=$75 nm, $T$=298 K, $M\approx$ 0.03 kg/mol (molar mass of air), and a total pressure $P\approx4\times10^{-8}$ mbar. As shown in Fig. \ref{fig:gasdamping}, the limit on quality factor imposed by the viscous damping model is at least an order of magnitude greater than the measured values of the fundamental modes. We also show here that second-order modes are more robust to gas damping than their first-order counterparts.

To further this investigation, an ExTorr XT100 quadrupole residual gas analyzer (RGA) outfitted with a hot cathode Bayard/Alpert (B/A) ion gauge was used to determine the relative gas composition in the science chamber (for species with masses < 100 amu) and to confirm the pressure inferred by monitoring the current of the test chamber's Varian Star-Cell 55 ion pump with the direct reading from the ExTorr's ion gauge. We found that the ion pump's inferred pressure ($P \approx4\times 10^{-8}$ mbar) compares favorably with the total pressure reading given by the B/A ($P = 4.1\times 10^{-8}$ mbar) and that the dominant gas species in the science chamber are atomic hydrogen, $H_2$, and nitrogen ($N_2$), all with partial pressures $p \approx6.7\times10^{-9}$ mbar. Other significant partial pressures present include water vapor, carbon dioxide, carbon-12, and small traces of 2-propanol and other organic solvents (used to clean the conflat flanges of the UHV chamber). Gas species in the mass spectrum were idenfied using the molecular weight search function of the NIST Chemistry WebBook, SRD 69.

\begin{figure}[ht]
    \centering
    \includegraphics[width=1.05\columnwidth]{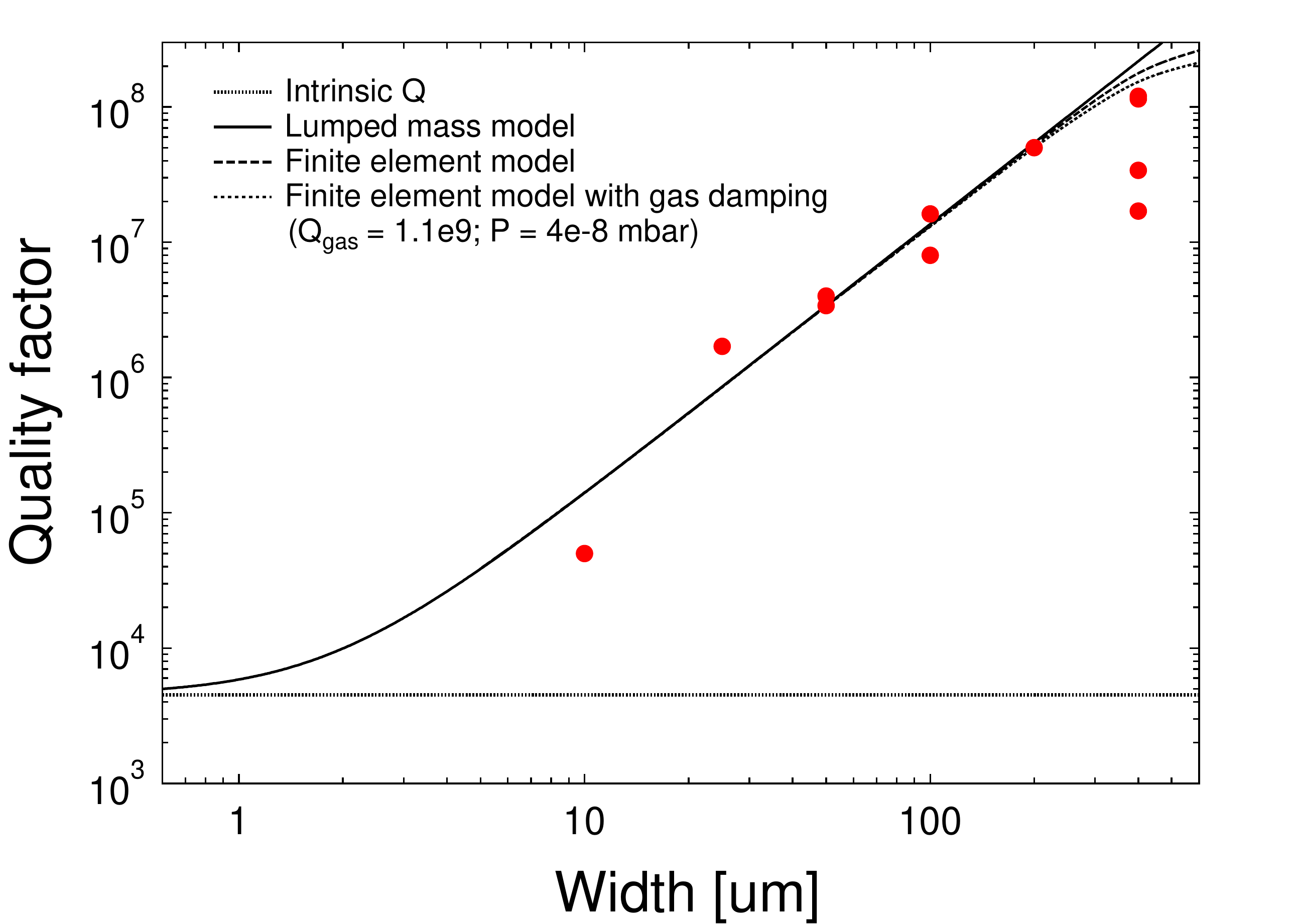}
    \caption{Quality factor versus width for the lumped mass model, finite element model, and finite element model including viscous damping effects. Red dots correspond to the fundamental torsion modes of a nanobeam.}
          \label{fig:FEMgasdamping}
\end{figure}

However, gas damping alone does not appear to be responsible for the reduction in quality factor that we observe (as compared with the lumped mass model for $w\geq$ 400 $\mu m$). When we include the finite element model, which accounts for the buckling instability discussed in the numerical modeling section, the culprit becomes apparent. We argue that both gas damping and buckling instabilities conspire to limit the ultimate quality factor of a nanobeam to $Q<1\times10^{9}$, shown in Fig. \ref{fig:FEMgasdamping}.

\newpage
\section{Quantum-limited optical lever: Theory}

In this section, we derive the shot-noise-limited angular resolution of an optical lever measurement, given by Eq. 2 in the main text:
\begin{equation}\label{eq:SNLSI}
    S_{\theta}^{\textrm{imp}} \ge \frac{1}{w_0^2} \frac{\hbar c \lambda}{8 P}.
\end{equation}

\begin{figure}[b]
\vspace{-3mm}
    \centering
    \includegraphics[width=0.6\columnwidth]{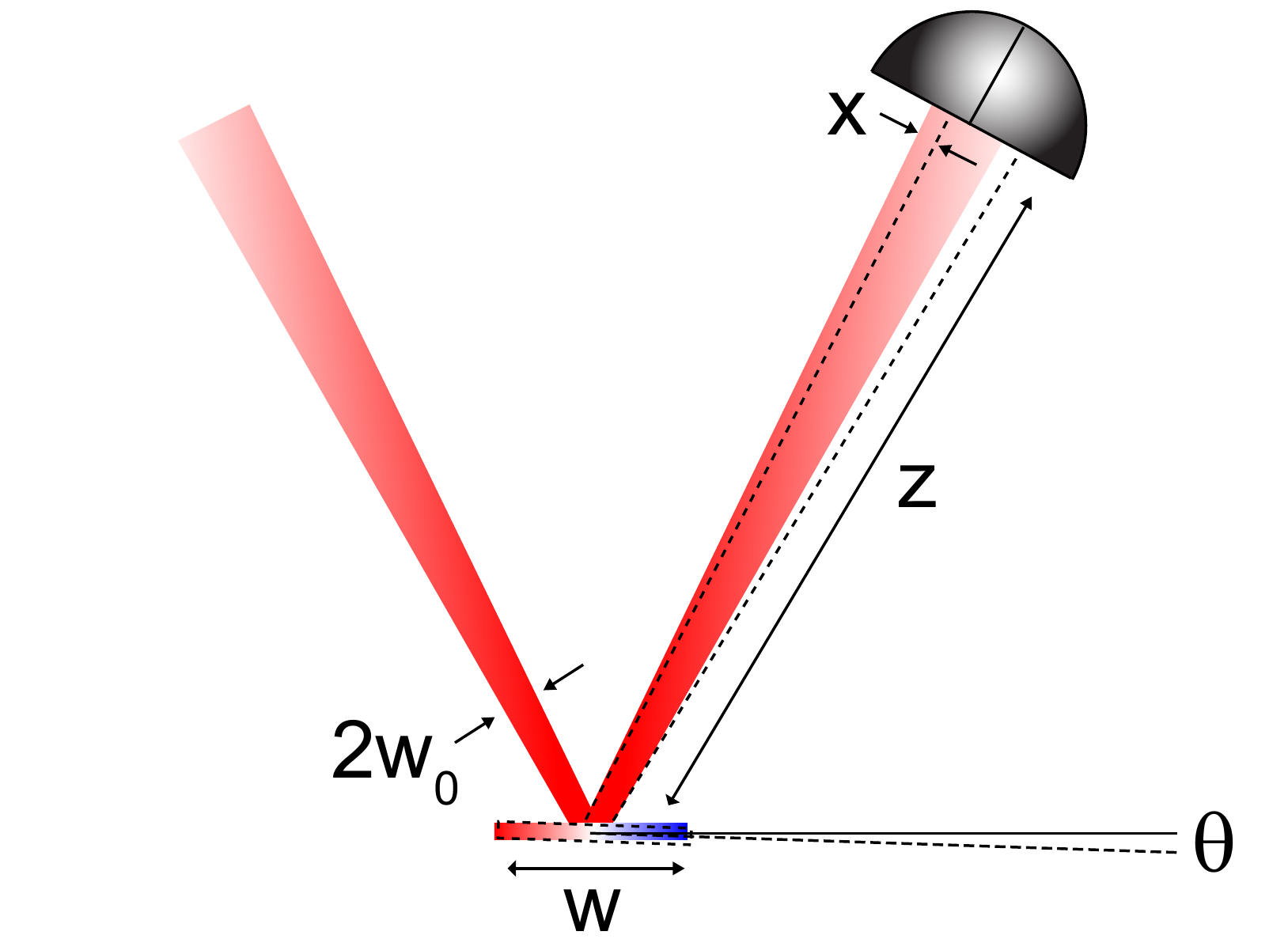}
    \caption{Schematic of optical lever measurement.  Angular displacement $\theta$ of a test surface produces a deflection $x = 2\theta z$ of a laser field on a split photodiode located a distance $z$ away.}
    \label{fig:OLtheory}
    \vspace{-5mm}
\end{figure}

As shown in Fig. \ref{fig:OLtheory}, an optical lever is formed by reflecting a laser beam from a test surface onto a split photodiode located a distance $z$ away. A small angular displacement of the surface $\theta$ results in a lateral displacement $x = 2\theta z$ of the laser beam on the photodiode. The resulting photocurrent is proportional to the split power difference
\begin{equation}
\label{eq:sensitivityIntegral}
\Delta P(x) = \int_{-\infty}^{x} I(x',y') dx'dy' - \int_{x}^{\infty} I(x',y') dx'dy'
\end{equation}
where $I(x',y')$ is the intensity of the laser field in the plane of the photodiode $x'-y'$ and $x'=0$ is the photodiode mid-line.

To obtain Eq. \ref{eq:SNLSI}, we let the laser be in a TEM$_{00}$ Guassian mode propagating normal to the photodiode with total power $P$, wavelength $\lambda$, waist size $w_0$, and waist location coinciding with the test surface\footnote{A careful analysis reveals this to be the optimal choice  \cite{putman1992}.}, such that
\begin{equation}
    I(x, y,z) = \frac{2 P}{\pi w(z)^2} \exp \left[ \frac{-2 (x^2 + y^2)}{w(z)^2} \right],
\end{equation}
where
\begin{equation}
w(z)=w_0\sqrt{1+z^2/z_0^2}
\end{equation}
is the radius of the beam on the photodiode and $z_0 = \pi w_0^2/\lambda$ is the beam's Rayleigh length.

Evaluating Eq. \ref{eq:sensitivityIntegral} yields the knife-edge signal
\begin{equation}
\Delta P (x,z) = P\;\t{Erf}\left[\frac{\sqrt{2}x}{w(z)}\right]
\end{equation}
which has a lateral displacement sensitivity of \cite{Treps_Quantum_2003}
\begin{equation}
\frac{\partial \Delta P}{\partial x}(x,z) = \frac{P}{w(z)}\sqrt{\frac{8}{\pi}} e^{-\frac{2x^2}{w(z)^2}}\approx \frac{ P}{w(z)}\sqrt{\frac{8}{\pi}}  
\end{equation}
and therefore an angular displacement sensitivity (referring to the displacement of the test surface, $\theta = x/(2z)$) of
\begin{equation}
\label{eq:OLSensitivity}
\frac{\partial \Delta P}{\partial \theta}(z) \approx \frac{2 P  z}{w(z)}\sqrt{\frac{8}{\pi}}
\end{equation}
in the limit of small displacements $x\ll w(z)\ll z$.

Eq. \ref{eq:OLSensitivity} appears to suggest that arbitrarily high sensitivity can be achieved by increasing the ``lever arm'' $z$; however, diffraction counterbalances the lever arm in the far field
\begin{equation}
    \frac{w(z)}{z}\xrightarrow[z\gg z_0]{} \frac{\lambda}{\pi w_0}\equiv\theta_0
\end{equation}
so that the sensitivity of the optical lever is bound above by
\begin{equation}\label{eq:OLsensitivity}
\frac{\partial \Delta P}{\partial \theta}\xrightarrow[z\gg z_0]{} \frac{2P }{\theta_0}\sqrt{\frac{8}{\pi}}
\end{equation}
where $\theta_0$ is the beam diffraction angle.

\begin{figure}[b]
    \includegraphics[width=0.95\columnwidth]{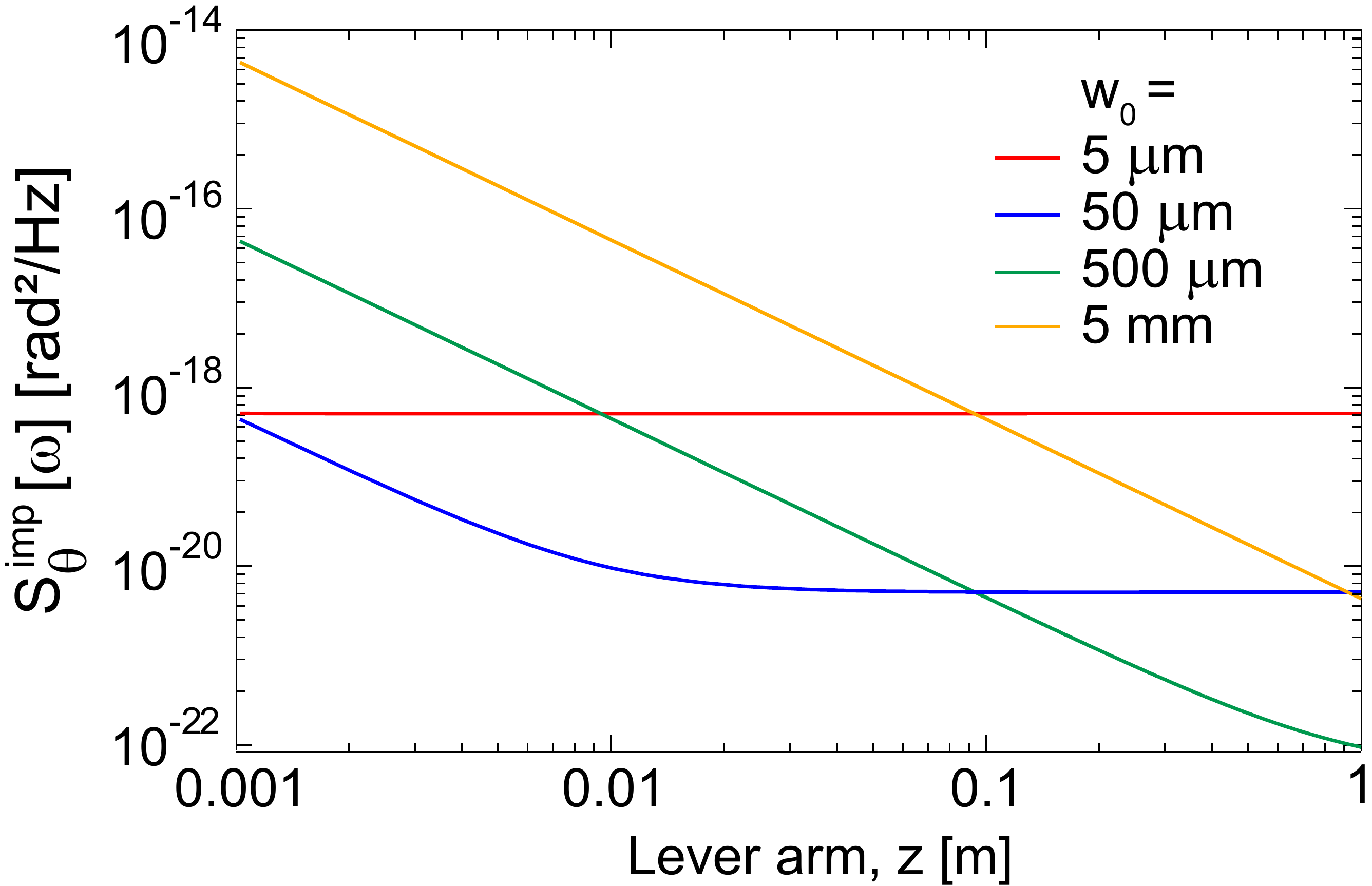}
    \caption{Shot noise limited resolution versus lever arm length for different waist sizes. Power is fixed at 1 mW.}
    \label{fig:SNL}
\end{figure}

To obtain Eq. 2 in the main text, we compare Eq. \ref{eq:OLsensitivity} to the fluctuations in optical power due to shot noise, here expressed as a single sided power spectral density (and assuming the laser is in a coherent state):
\begin{equation}
    S_{P}^{\textrm{shot}} = \frac{4\pi\hbar c}{\lambda}P.
\end{equation}
The angular displacement resolution (imprecision) is defined as the shot-noise-equivalent angular displacement
\begin{equation}\label{eq:OLshotresolution}
    S^{\t{imp}}_{\theta} = \frac{1}{\eta}\left(\frac{\partial \Delta P}{\partial \theta}\right)^{-2} S_{P}^{\textrm{shot}} \xrightarrow[z\gg z_0]{} \frac{1}{w_0^2} \frac{\hbar c \lambda}{8 \eta P}
\end{equation}
where $\eta\le 1$ is a parameter characterizing the measurement efficiency (including e.g. the photodetector efficiency). 

Equation \ref{eq:OLshotresolution} implies that optical levers with larger waist sizes achieve higher shot-noise resolution, at the expense of having to place the detector farther from the sample.  We visualize this in Fig. \ref{fig:SNL} by plotting $S^{\t{imp}}_{\theta}$ for optical levers with different waist sizes $w_0$ and fixed power $P=1$ mW, as a function of surface-detector distance ("lever arm") $z$.

\newpage
\section{Quantum-limited optical lever: Methods}
In this we section we give details on the high resolution optical lever measurement shown in Fig. 2 of the main text, including the experimental setup, calibration method, and evidence for the absense of photothermal heating, shot noise as the primary noise source, and off-resonant thermal noise (of higher order vibrational modes) as the technical noise limit.

\subsection{Experimental setup}

\begin{figure}[h]
   \centering
    \includegraphics[width=0.9\columnwidth]{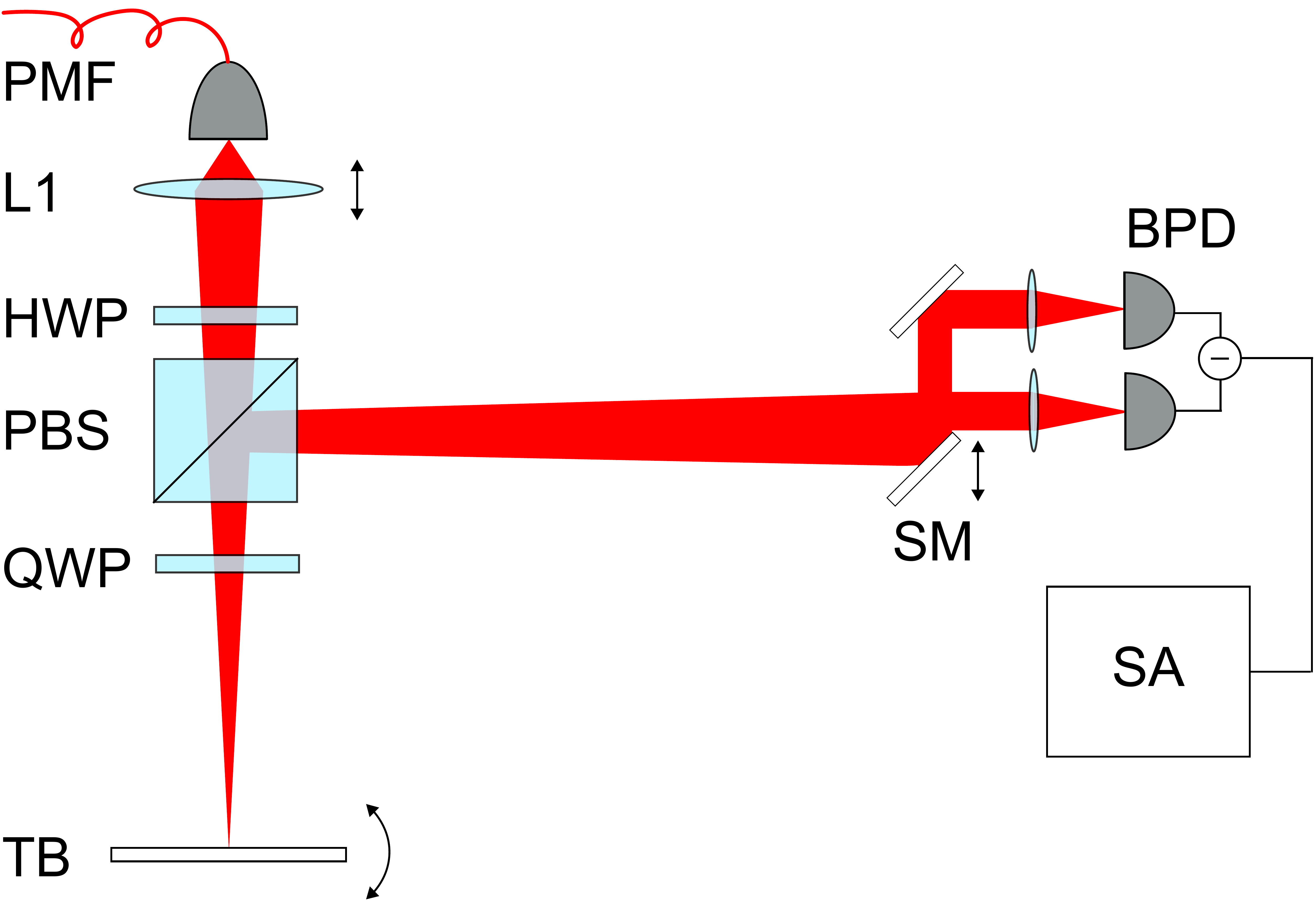}
    \caption{Optical lever setup for measurement in Fig. 2 of main text. PMF = polarization maintaining fiber, L1 = lens focusing light out of fiber, HWP = half-wave plate, PBS = polarizing beam splitter, QWP = quarter-wave plate, TB = torsion beam, SM = splitting mirror, BPD = balanced photodetector, SA = spectrum analyzer.}
    \label{fig:OLMeasurementSetup}
\vspace{-4mm}
\end{figure}

The experimental setup for Fig. 2 was the same as described for ringdown measurements, except the optical lever field was aligned normal to the sample in an autocolliminating configuration (using a PBS and a quarter waveplate as shown in Fig. \ref{fig:OLMeasurementSetup}) and the commercial split photodetector (Thorlabs PDQ80) was replaced by a custom split detector utilizing a straight-edged "splitting" mirror (SM) and a low noise balanced photodetector (Newport 1807). This setup allowed us to achieve shot noise limited performance with up to several milliwatts of optical power reflected from the sample.  

\subsection{Calibration}\label{sec:Calbration}

The measurement shown in Fig. 2 was calibrating by fitting the photocurrent spectrum, with detector noise substracted, to a noise model including the thermal motion of the torsion beam and imprecision noise:
\begin{equation}
    \label{eq:totalNoiseModel}
    S_{\theta}[\omega] = S_{\theta}^{\textrm{th}}[\omega] + S_{\theta}^{\textrm{imp}}[\omega].
\end{equation}
The imprecision noise is well approximated by white noise around resonance. For thermal noise we assume the angular displacement spectrum of a single mode, structurally damped oscillator \cite{gonzalez_brownian_1995}:
\begin{equation}
    \label{eq:TNM}
    S_{\theta}^{\textrm{th}}[\omega] = \frac{4 k_B T \omega_1 / (I_1 Q_1)} {(\omega_1^2 - \omega^2)^2 + \omega_1^2 \omega^2/Q_1^2},
\end{equation}
where $k_B$ is Boltzmann's constant, $T$ is the modal temperature (taken to be room temperature, $T = 292$ K), and $\omega_1$, $I_1$, and $Q_1$ are the angular resonance frequency effective moment of inertia, and quality factor of the fundamental torsion mode, respectively. The calibration factor is determined by knowledge of $\omega_1$ and $Q_1$, both of which we  measure, as well as $I_1$, which we determine using Eq. \ref{eq:momentOfInteria}. We only fit around resonance, so for simplicity we neglect the effects of structural damping and approximate the noise peak as a Lorentzian. When fitting, the only free parameter is the magnitude of the noise floor, $S_\theta^\t{imp}$.  Because the resolution the measured spectrum (0.1 Hz) is smaller than the linewidth of the mechancial oscillator (0.0005 Hz), \mbox{we mask the center of the noise peak when fitting.}

\begin{figure}[t!]
    \label{fig:calibrationComparison}
    \centering
    \includegraphics[width=0.95\columnwidth]{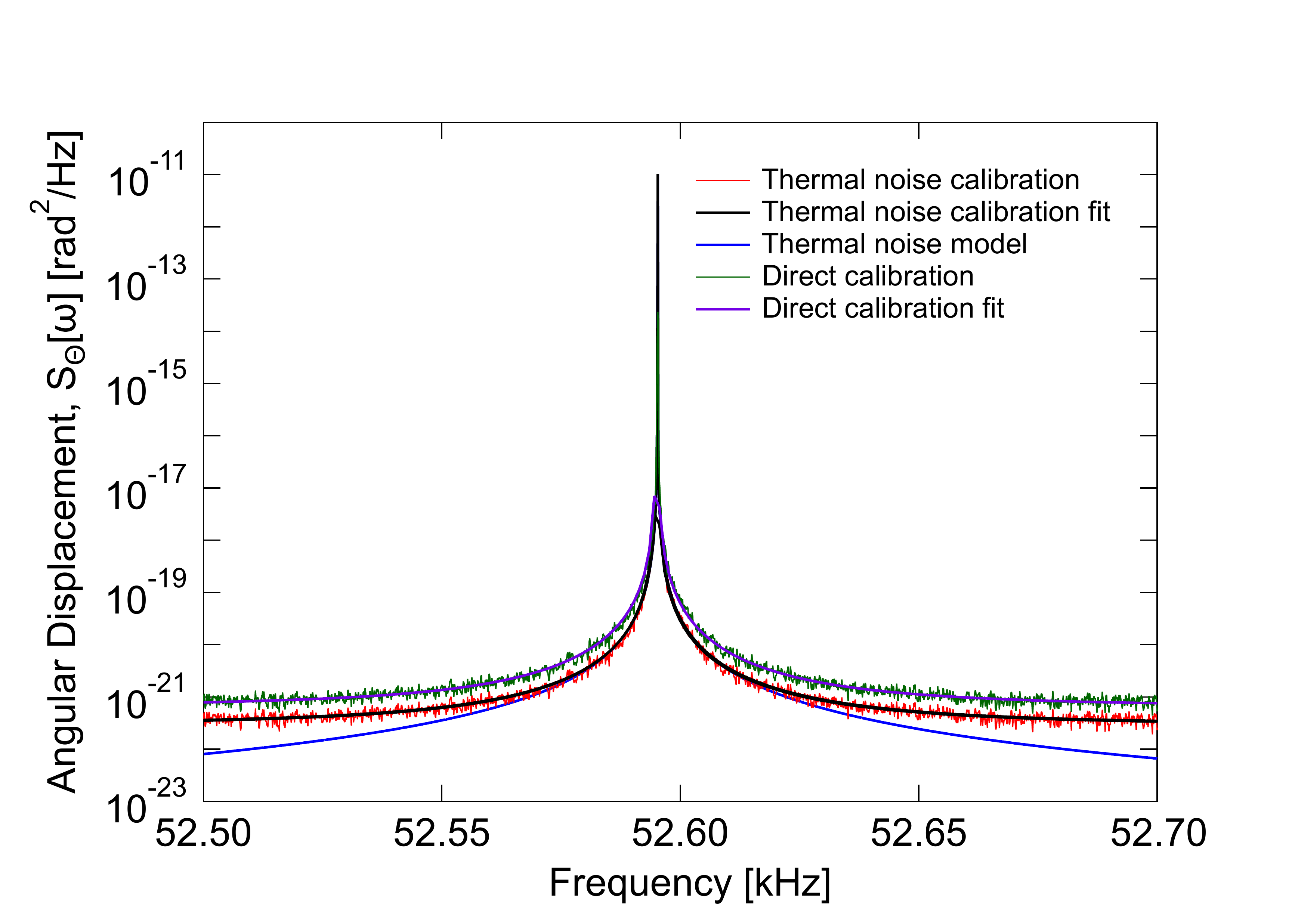}
    \caption{Comparison of calibration methods. Green curve: Direct calibration. Red curve: thermal noise calibration. Blue curve: thermal noise model without imprecision noise.}
\end{figure}

As a cross-check, in Fig. 10 we present an indepedent calibration of the angular dispacement spectrum by directly measuring the lateral position sensitivity of the split photodiode. The transduction factor from angular displacement to voltage $V$ at the output of the split photodetector is given by
\begin{equation}
        \frac{\partial V}{\partial \theta} = \frac{\partial V}{\partial x}\frac{\partial x }{\partial \theta}=2z\frac{\partial V}{\partial x},
\end{equation}
where $z$ is the distance from the beam to the photodetector (Fig. \ref{fig:OLtheory}) and  is the sensitivity to lateral beam displacements. We measured this quantity by translating the splitting mirror with a micrometer and recording the resulting voltage change. As shown in Fig. 10, the two calibration methods agree to within a factor of 2 in amplitude spectral density units. 

Finally, we re-emphasize, as stated in the main text, that the imprecision relative to the zero point motion, $S_\t{imp}/S_\t{zp}$ is independent of calibration method, and depends only the magnitude of the signal-to-noise ratio on resonance, viz.
\begin{equation}
    \frac{S_\theta^\t{th}}{S_\theta^\t{imp}}=2n_\t{th}\frac{S_\theta^\t{zp}}{S_\theta^\t{imp}}
\end{equation}
where $n_\t{th} = k_B T/\hbar\omega_n$ is the thermal mode occupation.  In our case $\omega_1 = 52.5$ kHz, $n_\t{th} = 1.2\times 10^8$, and $S_\t{th}/S_\t{imp} = 3.7\times 10^{10}$, from which we infer that $S_\t{zp}/S_\t{imp} = 140$.

\subsection{Beam waist characterization}
An important parameter for attaining the maximum angular displacement sensitivity is the beam waist size, which determines the diffraction-limited divergence angle of the beam. We characterize the beam waist by performing a knife-edge measurement at the focus. As a secondary check, we analyze the diffraction of the beam by performing knife edge measurements at different positions along the axis of propogation. 

\begin{figure}[t]
    \centering

    \includegraphics[width=0.85\columnwidth]{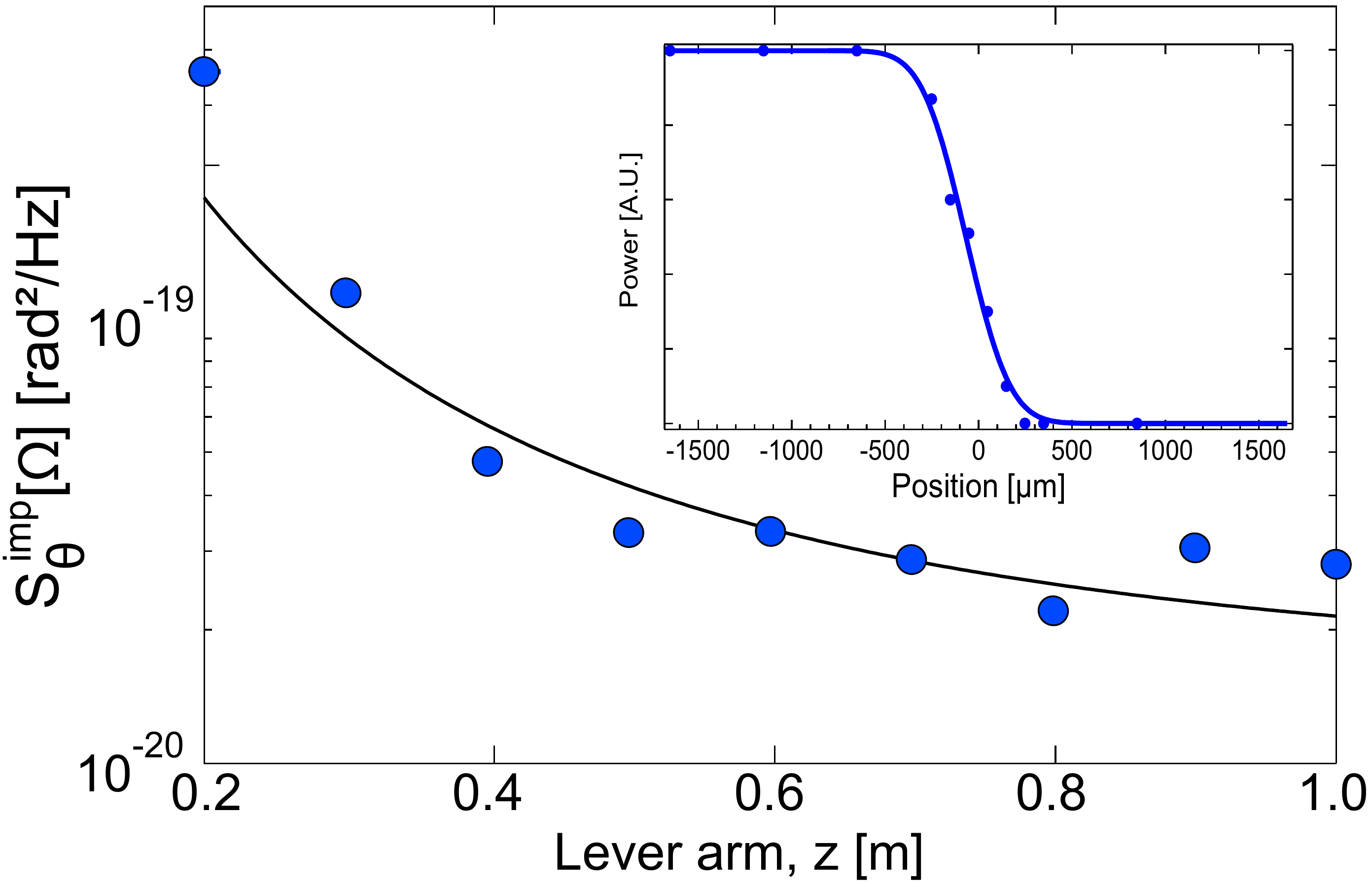}
    \caption{Imprecision versus detector-sample distance (``lever arm").  Black curve is a fit to Eq. \ref{eq:OLshotresolution} with efficiency $\eta$ and beam waist $w_0$ as free parameters, \textcolor{black}{yielding $w_0 = 413 \pm 216 \mu$m.}} 
    \label{fig:lengthSweep}
\end{figure}

Because the beam waist sets the length scale for the resolution of our measurement versus distance, we can utilize a scan of the sensitivity versus length to infer the beam waist as well. Figure \ref{fig:lengthSweep} shows a scan of the optical lever arm length versus the sensitivity, while fixing the power and beam waist size. The data was collected with a different split detection setup than used in the main text (Thorlabs PDQ80A), which contributed larger extraneous noise but made alignment easier, and utilized a larger beam size ($\sim 350 \mu$m).

\subsection{Shot noise model validation}
\begin{figure}[h]
    \centering
    \includegraphics[width=1\columnwidth]{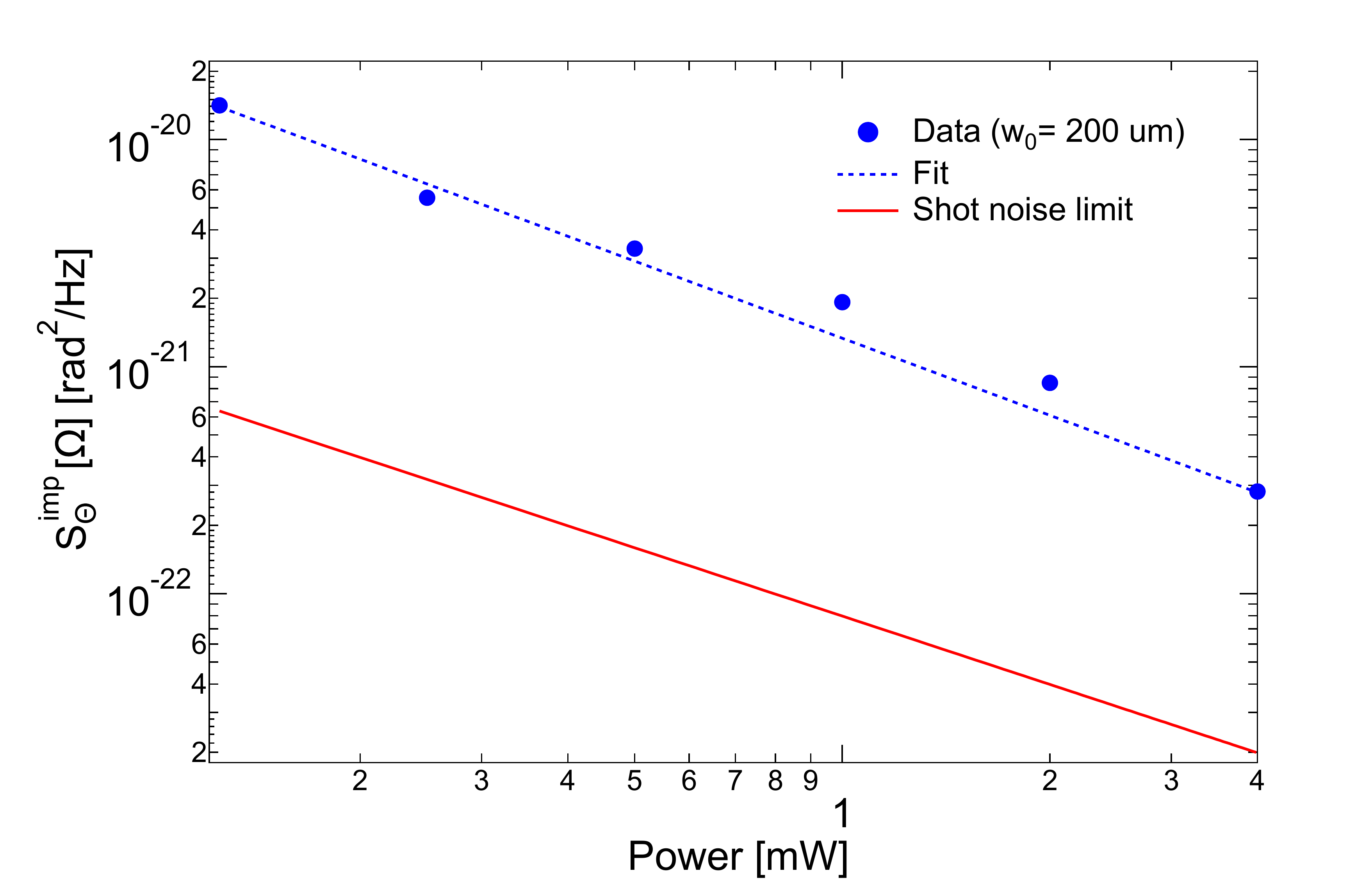}
    \caption{Imprecision versus power for two waist sizes. Points are measurements.  Solid lines are fits to Eq. \ref{eq:shotnoisefit}.  Dashed lines are the ideal shot noise limit given by Eq. \ref{eq:OLshotresolution}. Black line is an estimate of the extraneous noise due to off resonant thermal \mbox{motion (Sec. \ref{sec:otherModes}).}} 
    \label{fig:shotNoiseScaling}
\end{figure}

The noise floor in Fig. 2 of the main text is dominated by shot noise. To confirm this, as shown in Fig. \ref{fig:shotNoiseScaling}, \textcolor{black}{we recorded imprecision versus probe power $P$.} We then fit the data to a power law: 

\begin{equation}\label{eq:shotnoisefit}
    S_\theta^\t{imp} = \frac{\hbar c\lambda}{8 w_0^2 \eta} P^{b}.
\end{equation}
The fits yield $b = -1.12 \pm .09$ for a spot size of 200 $\mu$m, in good agreement with the expected scaling for shot noise ($b = -1$).

\subsection{Common mode rejection of classical intensity noise}

\begin{figure}[h]
    \centering
    \includegraphics[width=0.95\columnwidth]{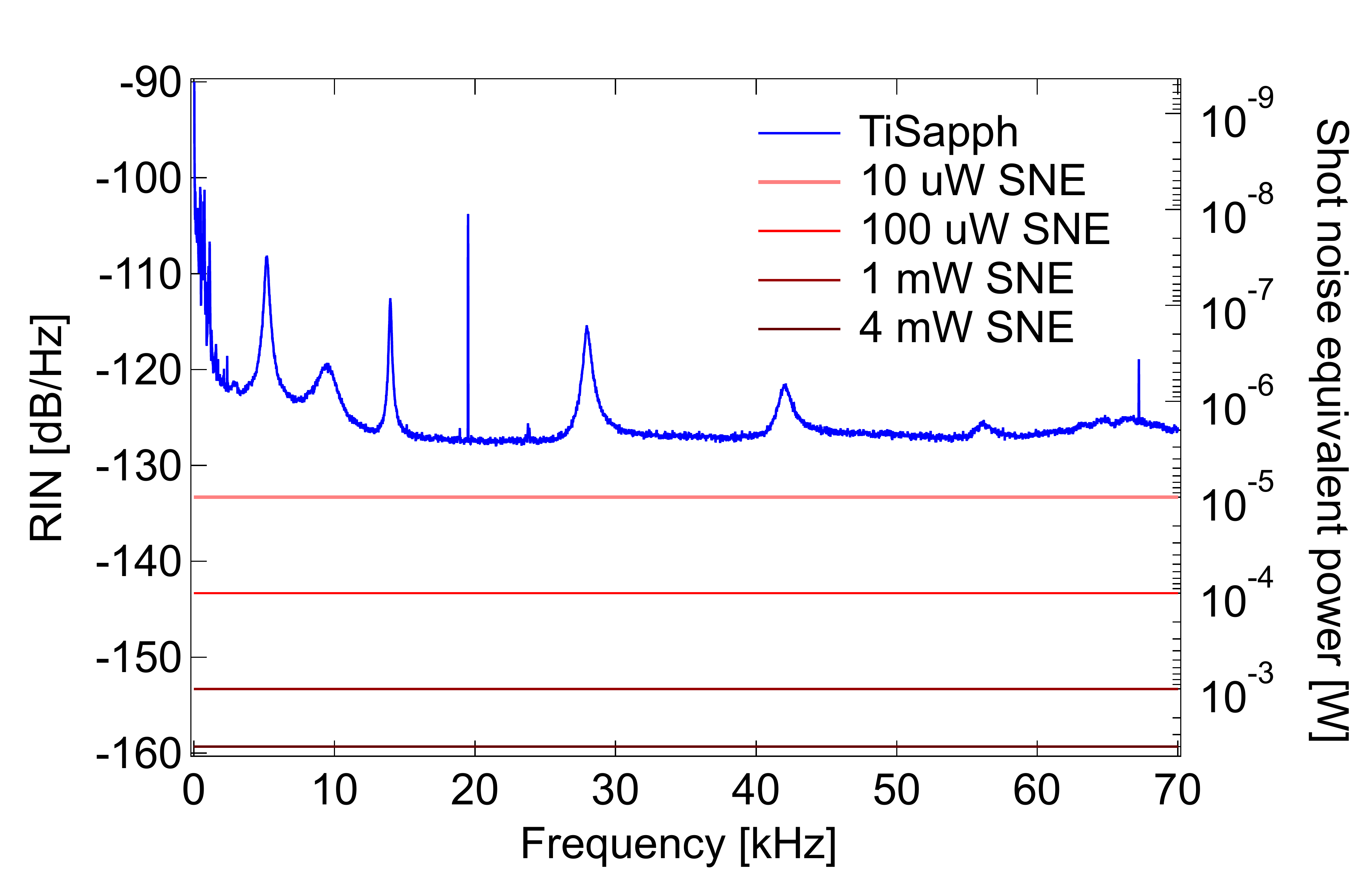}
    \caption{Relative intensity noise of the Ti-Sapphire laser used for Fig. 2 of the main text, compared to shot noise for $P=0.01-4$ mW.}
        \label{fig:laserIntensityNoise}
\end{figure}

Our measurements highlight that the optical lever technique naturally surpresses classical laser intensity noise; viz., the shot-noise (1/$P$) scaling in Fig. \ref{fig:shotNoiseScaling} was observed at powers for which the classical intensity noise of our Ti-Saphire laser, shown in Fig. \ref{fig:laserIntensityNoise}, overwhelmed shot noise by several orders of magnitude. 
Noise cancellation in excess of 30 dB was achieved by carefully translating the splitting mirror in \ref{fig:OLMeasurementSetup}, in order to balance the intensities on the two photodiodes.

\subsection{Multi-mode thermal noise spectra}

\label{sec:otherModes}
\begin{figure}[htbp]
    \centering
    \includegraphics[width=0.85\columnwidth]{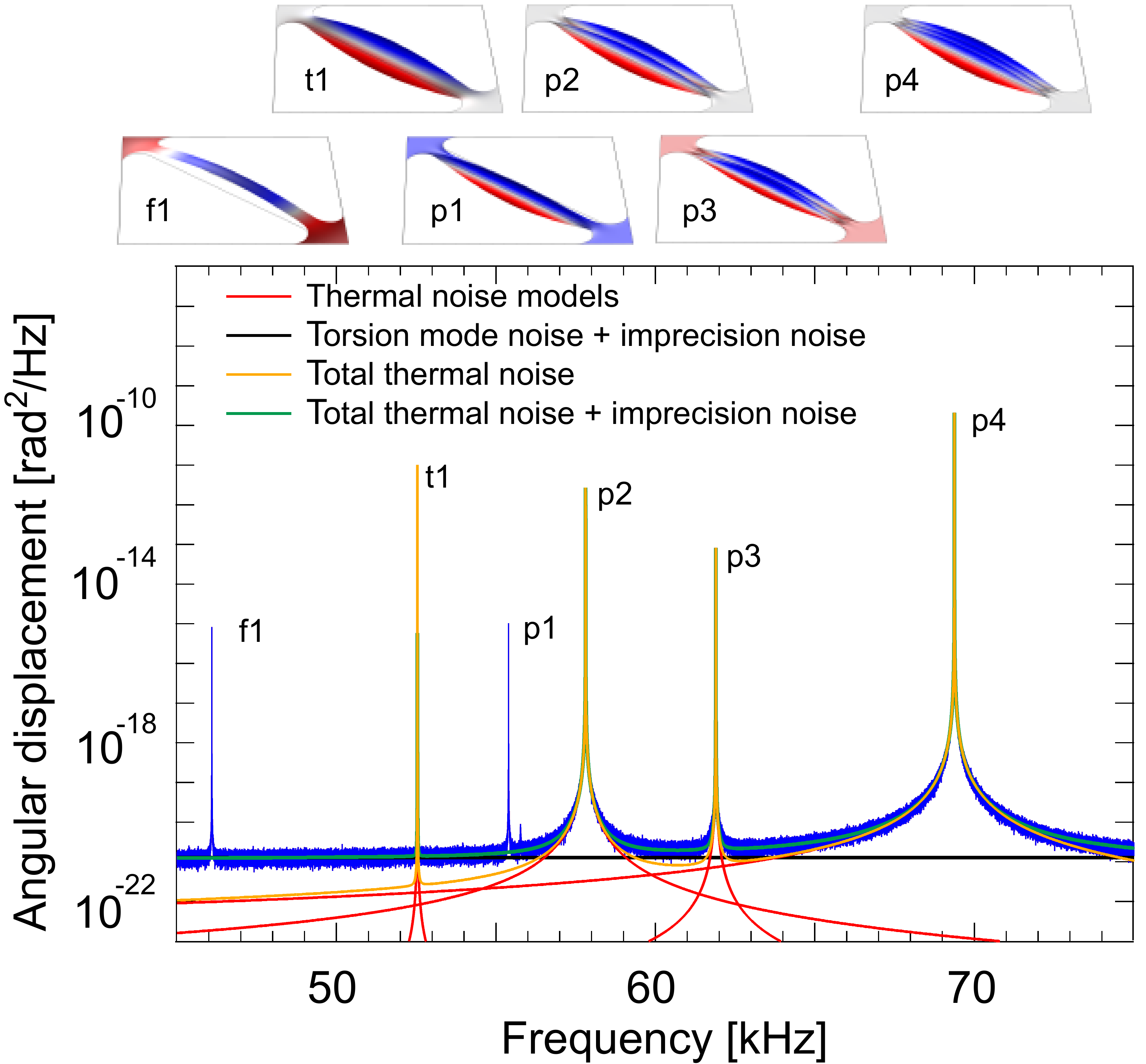}
    \caption{Broadband version optical lever measurement ($w = 400 \mu$m, $w_0 = 200\;\mu\t{m}$, P = 4 mW), with a slightly different alignment.  Thermal noise peaks of flexural (f), torsional (t), and potato-chip (p) modes are highlighted with corresponding modeshapes simulated in COMSOL.}
    \label{fig:higherordermodes}
\end{figure}

We have found that our optical lever measurements are sensitive to a broad variety of flexural and torsional modes, limited by the size and location of the laser spot on the beam. Fig. \ref{fig:higherordermodes} shows a measurement with a slightly different alignment than in the main text.  Owing to a slight misalightment, the lever is sensitive to both flexural modes and ``potato chip". Noteably, the off-resonant thermal noise from the potato chip modes at 57.5 kHz and 69 kHz is roughly 10\% (in power units) of the total noise in the viscinity of the fundamental torsional mode, at 52.5 kHz, corresponding to an extraneous noise (Eq. \eqref{eq:shotnoisefit}) of $S_\theta^\t{(ext)}= 2.7\times10^{-22} \textrm{ rad}^2/\textrm{Hz}$.

\newpage
\section{Acceleration sensitivity of a microtorsion pendulum}

\begin{figure}[b!]
    \includegraphics[width=0.9\columnwidth]{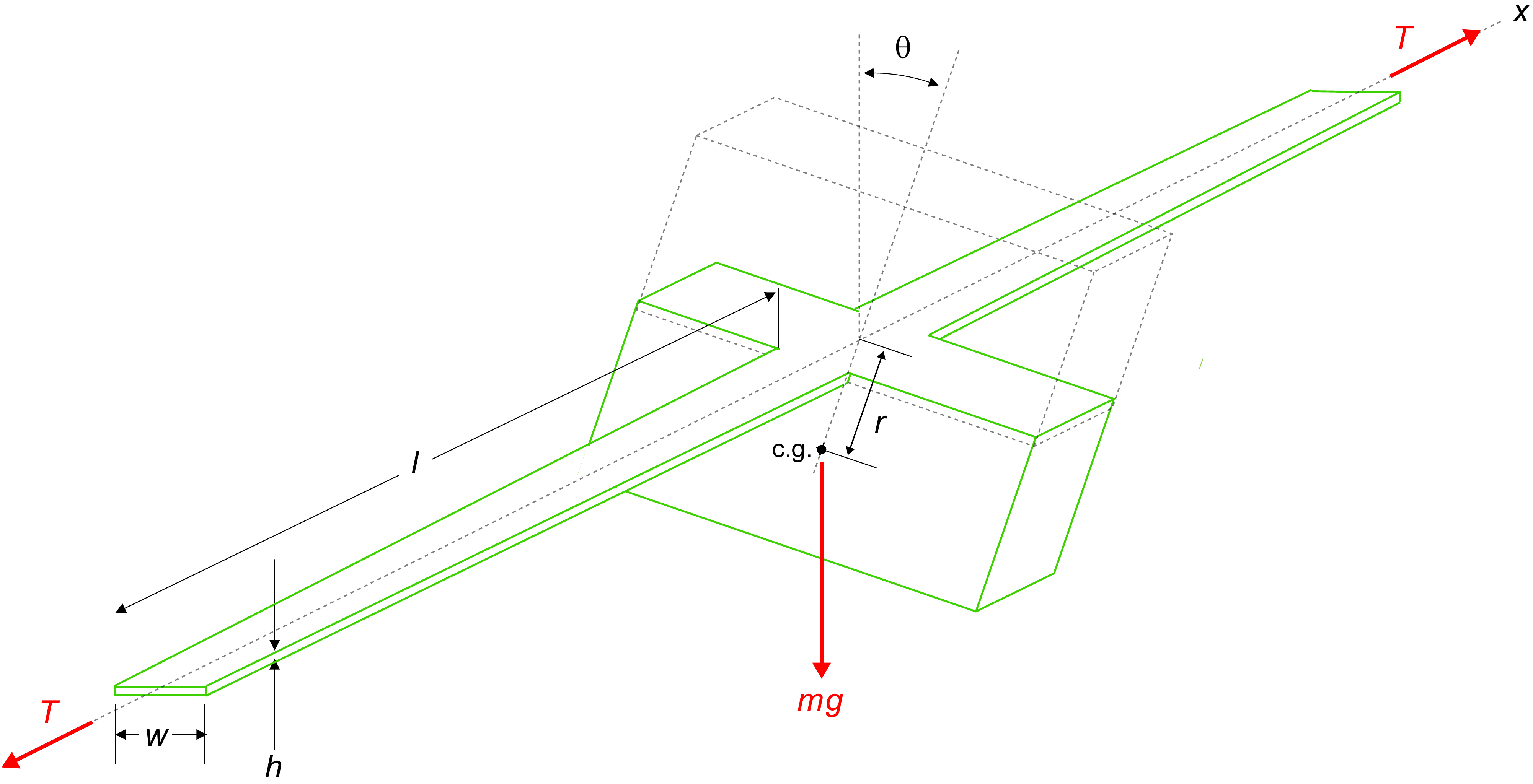}
    \caption{Schematic of torsion pendulum, formed by suspending a rectangular balance beam of mass $m$ from a ribbon-like torsion fiber under tension $T$.  Angular displacement of the balance beam $\theta$ is measured using an optical lever, described elsewhere.  The center of gravity (c.g.) of the balance beam is offset a distance $r$ from the rotation axis, resulting in a gravitational restoring torque, $\tau_g\approx mg r\theta $. }
    \label{fig:torsionbalance}
\end{figure}

In this section, we derive Eqs. 4-5 of the main text, describing the fundamental resonance frequency of a torsion balance with a tensile-stressed torsion fiber and the sensitivity of this frequency to local changes in the gravitational field strength.

Figure \ref{fig:torsionbalance} shows the basic geometry of the torsion balance with its torsion fiber oriented perpendicular to the earth's local gravitational field and a rectangular balance beam of mass $m$ free to oscillate, with moment of inertia $I$, about the fiber axis $xx$. The local acceleration of gravity $g$ acts on the balance beam with force $m g$ through its center of gravity (c.g.), offset from the pivot by a distance $r$.  Tilting the beam by a small angle $\theta$ produces a gravitational torque $\tau_g = mgr\sin(\theta)\approx m g r\theta$.

We assume the balance beam is symmetric about the torsion axis, so that its equilibrium tilt angle is zero.  The beam behaves in this case like a pendulum bob, executing simple harmonic motion (in the absense of damping or external torques) of the form
\begin{equation}\label{eq:torsionbalance1}
    I\ddot{\theta}+\kappa\theta = 0,
\end{equation}
where 
\begin{equation}\label{eq:torsionstiffnessG}
\kappa = \kappa_E + \kappa_\sigma \pm \kappa_g
\end{equation}
is the combined torsional stiffness do to elastic deformation $\kappa_E$, tensile stress $\kappa_\sigma$, and gravity
\begin{equation}
\kappa_g =  \frac{d\tau_g}{d\theta}= m g r,
\end{equation}
respectively.  The sign ($\pm$) of the gravitational term in Eq. \ref{eq:torsionstiffnessG} depends on the orientation of the pendulum, and is negative when the pendulum is inverted (c.g. above pivot).

Dividing Eq. \ref{eq:torsionbalance1} through by the moment of inertia, the squared frequency of the oscillator is given by
\begin{equation}\label{eq:torsionbalance2}
    \omega_\pm^2 = \frac{\kappa_E + \kappa_\sigma \pm \kappa_g}{I}
\end{equation}
which depends on gravity through the restoring torque $k_g$.

It follows that the sensitivity of the oscillator frequency to gravity in the non-inverted (Fig. \ref{fig:torsionbalance}) orientation is  
\begin{equation}\label{eq:torsionbalancesensitivity}
    \frac{d\omega_+}{d g}\approx\frac{\omega_+^2-\omega_-^2}{4 \omega_+ g_0},
\end{equation}
where $g_0 \approx 9.8\,\t{m}/\t{s}^2$ is the standard acceleration due to gravity on earth's surface.

We now seek an expression for the smallest detectable change in gravity, $\Delta g_\t{min}$. In the main text, we define $\Delta g_\t{min}$ as the shift in gravity needed to shift the oscillator frequency by it's full-width-half-maximum linewidth:
\begin{equation}
    \Delta\omega_+ = \frac{\omega_+}{Q_+},
\end{equation}
where $Q_+$ ($Q_-$) is the $Q$ factor of the oscillator in its non-inverted (inverted) configuration, ideally given by
\begin{equation}\label{eq:torsionbalanceQ}
Q_\pm = Q_0\left(1+\frac{k_\sigma\pm k_g}{k_E}\right) = Q_\mp\left(\frac{\omega_\mp}{\omega_\pm}\right)^2
\end{equation}
assuming that the gravitational stiffness is lossless (Eq. S1).

Thus we obtain Eq. 4 in the main text:
\begin{equation}\label{eq:gsensitivity}
    \Delta g_\t{min} \equiv \left(\frac{d\omega_+}{d g}\right)^{-1}\Delta\omega_+ = \frac{4 g_0}{Q_+}\frac{\omega_+^2}{\omega_+^2-\omega_-^2} = \frac{2 g_0}{Q_0}\frac{k_E}{k_g}
\end{equation}
where the latter equality assumes ideal dissipation dilution given by Eq. \ref{eq:torsionbalanceQ}.  Notably, $\Delta g_\t{min}$ is independent of the tensile stiffness $k_\sigma$.  This can be understood by observing that tension decreases the sensitivity of the oscillator (Eq. \ref{eq:torsionbalancesensitivity}) in the same proportion that it increases its $Q$ factor (Eq. \ref{eq:torsionbalanceQ}).

\subsection{Allan deviation measurement}

\begin{figure}[t!]
    \includegraphics[width=0.8\columnwidth]{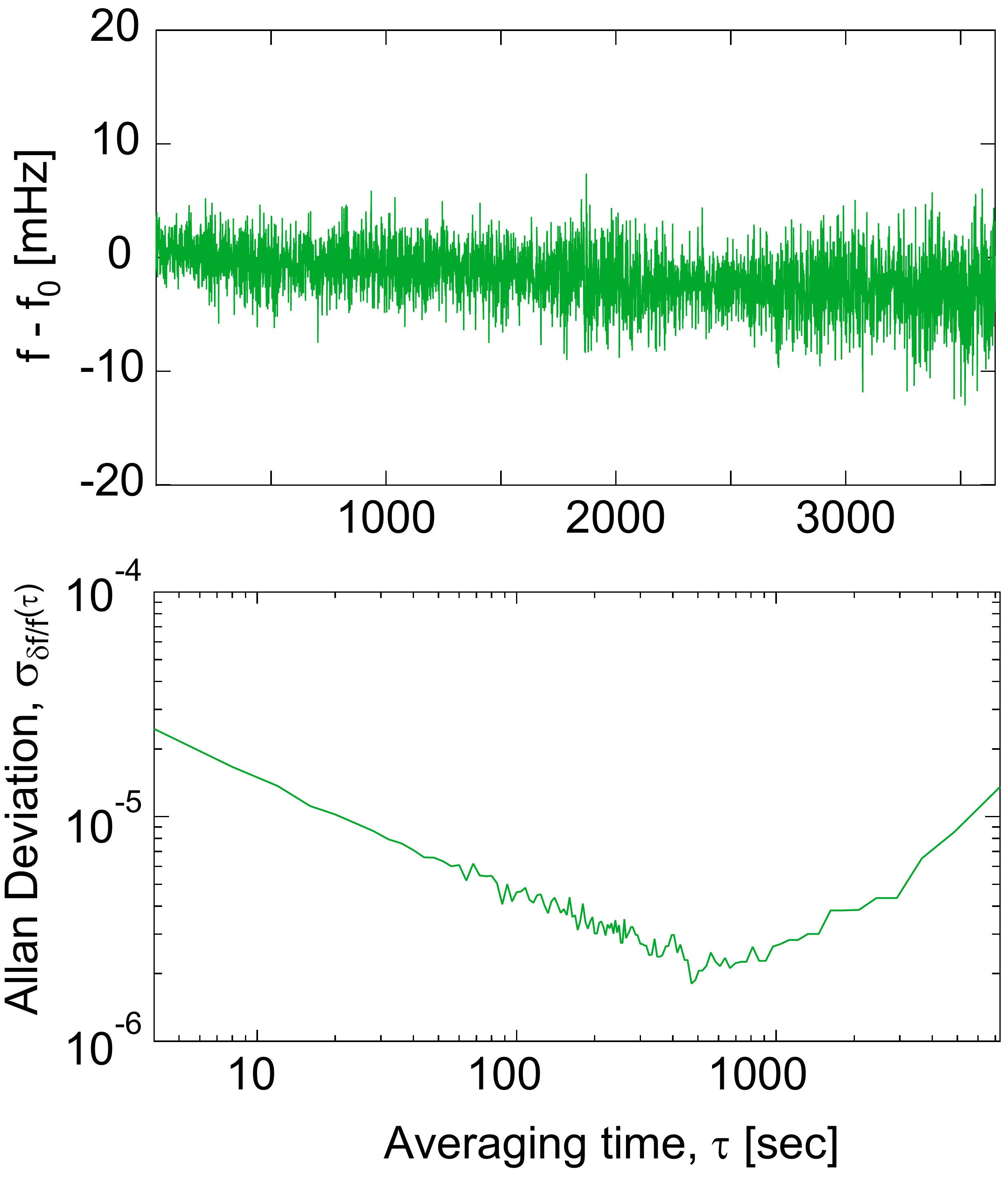}
    \caption{Frequency stability of microtorsion pendulum in Fig. 4 of main text. Above: frequency of pendulum (relatve to nominal start value of $f_0 \approx 35$ Hz) as a function of time.  Below:  Allan Deviation of frequency versus time measurement, normalized to $f_0$.}
    \label{fig:allandeviation}
\end{figure}

As mentioned in the main text, we carried out a long term measurement to assess the frequency stability of the 35 Hz micro-torsion pendula in Fig. 4.  The measurement is shown in Fig \ref{fig:allandeviation}.  He we tracked the free running resonance frequency of the non-inverted pendulum $f_+(t)=\omega_+(t)/(2\pi)$ by Fourier transforming a weak ($P\approx 5\mu$W) optical lever measurement over the course of a night.  The fractional Allan Deviation of the  time trace reaches a minimum value of $\sigma_{\delta f_+/f_+}\approx 2\times 10^{-6}$ at 600 seconds, corresponding to a gravitational acceleration uncertainty of $\sigma_{\delta f_+/f_+}\approx 8\times 10^{-6}$ according to Eq. \ref{eq:gsensitivity}.  We note that for this measurement, the temperature of the room was not stabilized, which may account for the observed $\sim \t{mHz}/\t{hr}$ frequency drift.

\newpage

\bibliographystyle{apsrev4-1}
\bibliography{ref2}